\documentclass[11pt]{article}
\usepackage{amsmath}
\usepackage{color}
\usepackage{mathrsfs}
\usepackage{amssymb}
\usepackage{amsthm}
\usepackage{epstopdf}
\usepackage{graphicx}
\usepackage{mathtools}
\usepackage[usenames,dvipsnames, table]{xcolor}
\usepackage[hang]{caption}
\usepackage{hyperref}
\usepackage{curve2e}

\usepackage{oldgerm}  
\usepackage[yyyymmdd,hhmmss]{datetime}

\usepackage{mathabx} 
\usepackage{dbnsymb}

\def\bee{\begin{enumerate}}\def\eee{\end{enumerate}}
\def\bei{\begin{itemize}}\def\eei{\end{itemize}}
\newcommand{\nco}{\newcommand}
\def\R{\mathbb{R}}\def\C{\mathbb{C}}
\nco{\red}{\color{red}}
\nco{\blue}{\color{blue}}
\nco{\green}{\color{green}}
\nco{\cyan}{\color{cyan}}
\nco{\brown}{\color{Magenta}}

\nco{\magenta}{\color{magenta}}

\nco{\violet}{\color{violet}}
\nco{\olive}{\color{Emerald}}
\nco{\orange}{\color{orange}}
\nco{\redend}{\normalcolor}
\nco{\blueend}{\normalcolor}

\def\tr{{\rm tr}\,}

\def\ommit#1{{}}
\def\({\left(}
\def\){\right)}
\def\ie{{\it i.e.,\/}\ }
\def\ie{{\rm i.e.,\/}\ }

\definecolor{cb}{rgb}{.8,.5,0}

\nco{\rnc}{\renewcommand}
\rnc{\title}[1]{{\Large\bf\mbox{}\\\medskip#1\bigskip\medskip\\}}
\rnc{\author}[1]{{\large #1\smallskip\\}}
\nco{\address}[1]{{\em #1\medskip\\}}

\nco{\bun}{{\bf 1}}
\def\be{\begin{equation}}\def\ee{\end{equation}}
\def\bea{\begin{eqnarray}}\def\eea{\end{eqnarray}}
\def\bee{\begin{enumerate}}\def\eee{\end{enumerate}}
\def\bei{\begin{itemize}}\def\eei{\end{itemize}}

\def\ommit#1{{}}

\def\SU{{\rm SU}}

\def\tr{{\rm tr\, }}

\def\eq=#1{\buildrel #1 \over{=}}

\def\N{\mathbb{N}}
\def\Z{\mathbb{Z}}

\def\R{\mathbb{R}}
\def\C{\mathbb{C}}

\newtheorem{definition}{Definition}
\newtheorem{proposition}{Proposition}

 \DeclareMathOperator{\dimension}{dim}
\DeclareMathOperator{\tri}{tri}
\DeclareMathOperator{\tet}{tet}
\DeclareMathOperator{\gon}{gon}
\DeclareMathOperator{\hed}{hed}
\DeclareMathOperator{\M}{M}
\DeclareMathOperator{\TET}{TET}
\DeclareMathOperator{\SixJSymbol}{SixJSymbol}

\begin{document}
\begin{titlepage}
\begin{center}
\title{About integer-valued variants of the theta and $6j$ symbols}
\medskip
\author{Robert Coquereaux} 
\address{Aix Marseille Univ, Universit\'e de Toulon, CNRS, CPT, Marseille, France\\
Centre de Physique Th\'eorique}
\bigskip
\begin{abstract}
These notes contain essentially a rewriting of several  properties of two well-known quantities, the so-called theta symbol (or triangular symbol),  which is rational, and the $6j$ symbol, which is usually irrational, in terms of two related integer-valued functions called $\gon$ and $\tet$.  
Existence of these related integer-valued avatars, sharing most essential properties with their more popular partners, although a known fact, is often overlooked.
The properties of $\gon$ and $\tet$ are easier to obtain, or to formulate, than those of the corresponding theta and $6j$ symbols, both in the classical and  quantum situations. 
Their evaluation is also simpler (the paper displays a number of explicit formulae and evaluation procedures that may speed up some computer programs).
 These two integer-valued functions are unusual, in that their properties do not appear to be often discussed in the literature, but their features reflect those of related real-valued functions discussed in many places.
Some of the properties that we shall discuss seem however to be new, in particular {  several relations between the function $\gon$} and the inverse Hilbert matrices.

\end{abstract}
\end{center}
\end{titlepage}
\section{Introduction}

This set of notes started as a a pedagogical exercise, or as a self-directed learning activity. 
The purpose was to present the notions underlying the definitions and properties of several well-known functions of mathematics and theoretical physics, namely the so-called triangular (or trihedral, or theta) symbols and  $6j$ symbols, in very simple terms.

These functions are used by several communities working in often quite distinct fields, for instance elementary quantum physics, chemistry, spin networks and spin foams, particle physics, quantum gravity,
topological field theory, geometry of 3-manifolds, representation theory of semi-simple Lie groups, weak Hopf algebras, category theory, and even quantum computing, to mention just a few.
Quite often the various definitions of these functions, and the writing of their properties, appears as rather involved, in the sense that they require some familiarity with specific scientific domains.
The terminology itself is not fully standardized because several objects bearing the same name often differ by normalization prescriptions or have a meaning that has not be stable along the years.

The main observation is the following: there exists a variant of the theta symbols (triangular symbols) that makes them integer-valued, and there is also a variant of the $6j$ symbols that makes them integer-valued.
The integrality of these two variants comes from the fact that they can be written as multinomial coefficients (or sums of multinomials).
The ``usual'' theta and $6j$ symbols are related to their respective integer-valued variants by normalization coefficients that have an elementary geometrical interpretation.
This fact is certainly known by experts (or aficionados) but it is often overlooked, and the absence of such a description in the literature ---with some rare exceptions, like~\cite{GaroufalidisEtAl}--- is quite surprising.

We therefore decided to write a kind of compendium of formulae involving  these integer-valued avatars of the theta and $6j$ symbols.
The paper does not contains many proofs because most formulae displayed in the following pages come from well-documented properties of the theta and $6j$ symbols but they are usually simpler and look more natural  when written in terms of their integer-valued variants.
We also decided to give a name to these two integer-valued functions:  in order to avoid the notation $\tri$ (or $\theta$) for the variant of the triangular function, we decided to call it $\gon$ (also because it can be generalized to polygons), and to call $\tet$ the integer-valued variant of the $6j$-symbol. As already mentioned, most formulae to be found in the next pages are nothing but a re-writing of known relations, however we shall discover a few others along the way; in particular we shall discover unexpected relations with the coefficients of inverse Hilbert matrices (see sec.~\ref{sec:Hilbert}).
A double triangle identity discussed in sec.~\ref{sec:doubletriangleidentity} does not seem to be much discussed in other places either.

It turns out that the $q$-number generalization of the two integer-valued functions $\gon$ and $\tet$ is very simple to obtain, and we shall describe them as well (the ``classical'' $6j$ symbols are used in elementary quantum physics, but their quantum analog are not !).
The first part of this paper is devoted to the definition and study of the ``classical'' functions $\gon$ and $\tet$ whereas the second part (more sketchy)  is devoted to their ``quantum'' counterpart.
Several comments are gathered in the last section (Miscellaneous) where the reader will in particular find a very brief summary of the spin network formalism ---that we shall not use in the main body of the text, but which is related to our previous discussions.

There is a huge literature on the mathematics of the triangular and $6j$ symbols, most of the corresponding results  are rather old and now belong to the folklore. 
{  Selecting references is not an easy task, some being more appropriate than others for readers with a specific background;
 we shall nevertheless mention here a few classical textbooks.
 General or historical ones belong to the physics literature, we can mention  \cite{Hecht}, \cite{Messiah}.} 
The mathematically oriented reader having particular applications in mind will find useful informations in  \cite{Carter:6J}, \cite{KauffmanLins}, \cite{KlimykSchmudgen},  \cite{Roberts:tetrahedron}.
One can also look at the appropriate sections of Wikipedia.
{  Finally we included a short appendix that briefly recalls the historical motivation behind the definitions of various families of coefficients (Clebsch-Gordan coefficients, $3j$ symbols, $6j$ symbols) and where we display a few relations that are standard, 
but not every reader is expected to be familiar with the subject. We shall sometimes refer to this appendix for notational purposes and when we need to make contact between the functions $\gon$ or $\tet$ discussed in the main body of the text and more traditional quantities.}

\section{Classical version}
 
\subsection{The classical gon function}

\subsubsection{Admissible triplets}
\label{admissibletriple}
Let $(a,b,c)$ be a triplet of non-negative integers.  We first assume that this triplet is {\sl triangular}, meaning that it obeys triangular inequalities: the  numbers $a,b,c$ are such that  $a+b-c\geq 0$, $b+c-a\geq 0$ and $c+a-b\geq 0$.
These integers can therefore serve as the side lengths of an integer triangle (possibly degenerate, either because one edge is zero or because one of the  triangular inequalities is an equality).
We also assume that the perimeter $a+b+c$ is even. A triplet of non-negative integers $a,b,c$ obeying the above two constraints is called {\sl admissible}. 

Notice that $a+b-c = (a+b+c)-2c$ is a difference of two even integers; hence $a+b-c$  is an even integer, and this is true as well for $b+c-a$ and $c+a-b$.
One can therefore introduce three new non-negative integers $m,n,p$ as follows:  \[m = \frac{1}{2} (c+a-b), \quad n=\frac{1}{2} (a+b-c), \quad p=\frac{1}{2} (b+c-a).\]
Then $a=m+n$, $b=n+p$, $c=p+m$.  The semiperimeter $\sigma$ of the triangle $(a,b,c)$ reads  $$\sigma=(a+b+c)/2=m+n+p,$$ and we have $m=\sigma-b$, $n=\sigma-c$,  $p=\sigma-a$.
Notice that $(a,b,c)$  admissible implies $m~\geq~0$, $n\geq 0$, $p\geq 0$, but this does not imply that $(m,n,p)$ should be triangular: $m, n, p$  can be arbitrary non-negative integers.
It is clear from the above that there is a one to one correspondance between the triplets $(a,b,c)$ and the triplets $(m,n,p)$. We find convenient to refer to the first as {\sl external} variables, and to the next as  {\sl internal} variables;
the external variables label edges of triangles, whereas the internal variables should be thought as labels for the segments displayed on fig.~\ref{oblade_su2}.

It is sometimes useful --- and anyway traditional in physics --- to introduce the spin variables $j_1=a/2$, $j_2=b/2$, $j_3=c/2$.  One has $m=j_3+j_1-j_2$, $n=j_1+j_2-j_3$, $p=j_2+j_3-j_1$.
These so-called spin variables  $j_1, j_2, j_3$ are either integers or half-integers. Notice that $m+n+p=j_1+j_2+j_3$.
  
One can interpret also $a$, $b$, and $c$ (twice the spin variables)  as the components of the highest weights defining three irreducible representations (irreps) of $SU(2)$ in the basis of fundamental weights --- since $SU(2)$ has only one fundamental weight this basis has only one element.
We shall usually denote the irreducible representations themselves by their highest weight.
The dimension of the irrep with  highest weight component equal to $x=2j$ is $\text{dim}(x)=x+1=2j+1$.
In this interpretation, a triplet of representations $(a,b,c)$ is admissible if $c$ (for instance) enters the decomposition of  the tensor product $a\otimes b$ as a sum of  irreducible representations, equivalently, if $a\otimes b\otimes c$ contains the trivial representation, \ie if and only if $c \in \{|a-b|, |a-b|+2,\ldots, a+b-2, a+b\}$. {  As it will appear many times, we give a name to the previous set: {  for non-negative integers $u,v$, we call 
\begin{equation}
\bigotimes[u,v]=\{|u-v|,|u-v|+2,\ldots, u+v-2,u+v \}
\label{bigotimesset}
\end{equation}}}
  
\subsubsection{The elementary $\gon$ function (triangular function)}  
\label{gon:definition}
For an admissible triplet $(a,b,c)$, the function $\gon(a,b,c)$ is defined in terms of the variables $m, n, p$, introduced previously, as a particular multinomial coefficient:
{\definition
 \begin{equation}{\gon}(a,b,c)=\text{Multinomial}(m,n,p,1)=\frac{(m+n+p+1)!}{m! n! p!}    \label{gondefinition}   \end{equation}}
Explicitly,
 \begin{equation}{\gon}(a,b,c)=\frac{\left(\frac{1}{2} (a+b+c)+1\right)!}{\left(\frac{1}{2} (a+b-c)\right)! \left(\frac{1}{2} (a-b+c)\right)! \left(\frac{1}{2} (-a+b+c)\right)!}\label{gonfromfactorial}\end{equation}

Since $\text{Multinomial}(m,n,p,1)$, also written $\M(m,n,p,1)$ for short, is a multinomial coefficient,  the value of $\gon(a,b,c)$ for an admissible triplet is  automatically an integer. 
By construction, the function $\gon$
is symmetric in its three arguments. Using the semi-perimeter variable $\sigma$ introduced previously and the fact that $M(m,n,p,1)=(\sigma+1)M(m,n,p)$ we can also write $\gon$ as follows: 
 \begin{equation}
\gon(a,b,c)=(\sigma +1) \, M(\sigma-a, \sigma-b, \sigma-c)
\end{equation}
where $M(\sigma-a, \sigma-b, \sigma-c)$ is the multinomial coefficient $\frac{\sigma !}{(\sigma -a)! (\sigma -b)! (\sigma -c)!}$.

\smallskip

The reader familiar with the theory of spin networks will have recognized that the function $\gon$ is, up to sign, equal to one of the avatars (one of the evaluations) of the so called theta graph, usually denoted ${\ThetaGraph}$.
One should however be cautious because there are several evaluations prescriptions for the evaluation of spin networks, differing by normalization factors.  We shall  
give a brief account of this theory in sec.~\ref{sec:classicalspinnetworks}  (see also \ref{KauffmanTheta}) and compare the different prescription types.
Among them, one may be called the ``integer evaluation prescription'', precisely because the evaluation gives an integer. In particular the evaluation of the ${\ThetaGraph}$ graph with edges $a,b,c$, using this particular prescription, is denoted ${\ThetaGraph}_\Z(a,b,c)$
and it is enough, for the moment, to mention that $${\ThetaGraph}_\Z(a,b,c) =  (-1)^\sigma \, \gon(a,b,c).$$

{We extend the definition} of the function $\gon(a,b,c)$ by declaring that it vanishes if its three arguments do not build an admissible triplet of non-negative integers (this is also what is done for ${\ThetaGraph}$ in the theory of spin networks).
Another possibility, given three arbitrary real positive numbers $a,b,c$, could be to introduce the same variables $m,n,p$ as before  but use Gamma functions rather than factorials in the definition (\ref{gondefinition})  of $\gon$;
even more generally one could think of using the multivariate Euler Beta function $B$ with three arguments,  $B(m, n, p) = \Gamma(m) \Gamma(n) \Gamma(p)/\Gamma(m + n + p)$, and consider the function defined as
$1/{(\left(\sigma+2\right) B\left(\sigma-a+1,\sigma-b+1,\sigma-c+1\right))}$.
 However, we shall not explore such possibilities here.  

\smallskip
In section (\ref{generalpolygon}) we shall generalize the notion of admissibility and extend the definition of the $\gon$ function  in such a way that it admits an arbitrary number of compatible arguments.

\subsubsection{Particular cases and elementary relations}  

$\bullet$ In the degenerate case $a=b$, $c=0$, one has $m=p=0$ and $n=a=b$, therefore $gon(a,a,0)= n+1=a+1$, which is equal to $\text{dim}(a)$ if the integer $a$ is interpreted as the component of the highest weight  labelling an irreducible representations of $SU(2)$.\\
$\bullet$ From the definition of $\gon$ and taking $a=b=c$ even, one finds immediately $m=n=p=a/2$ and  $\gon(a,a,a)=M(n,n,n,1)=(3n+1)!/n!^3$, which is the OEIS sequence A331322.\\
$\bullet$ From the definition \ref{gondefinition},  simple manipulations lead to the following expressions (involving binomial coefficients) that do not look symmetric in the variables $a,b,c$ but show, incidentally, that ${\gon}(a,b,c)$ is divisible by $(a+1)$, by $(b+1)$, and by $(c+1)$:
  {\scriptsize
  \begin{equation}
  \label{gonexpression}
    \begin{split}
 {\gon}(a,b,c)=&\M(m,n,p,1) =\\
 (p+m+1) \binom{m+n+p+1}{n} \binom{p+m}{p} =& (m+n+1) \binom{m+n+p+1}{p} \binom{m+n}{m} =  (n+p+1) \binom{m+n+p+1}{m} \binom{n+p}{n},  \end{split}\end{equation}} \\
$\bullet$ From the Pascal's simplex relation  {\footnotesize$\M(m - 1, n, p, 1)+ \M(m, n - 1, p, 1) +  \M(m, n, p - 1,  1) =  (m + n + p)/(m + n + p + 1) \M(m, n, p, 1)$} one obtains immediately
{\footnotesize\[ {\gon(a-1,b,c-1)}+{\gon(a-1,b-1,c)}+ {\gon(a,b-1,c-1)}=\frac{\frac{1}{2} (a+b+c)}{\frac{1}{2} (a+b+c)+1} \times {\gon(a,b,c)}\]}\\
$\bullet$ From the identity {\footnotesize $B(m + 1, n, p) + B(m, n + 1, p) + B(m, n, p + 1) = B(m, n, p)$} for the multivariate Euler Beta function one also obtains:
{\scriptsize\[ \frac{1}{\gon(a+1,b,c+1)}+\frac{1}{\gon(a+1,b+1,c)}+\frac{1}{\gon(a,b+1,c+1)}=\frac{\frac{1}{2} (a+b+c)+3}{\frac{1}{2} (a+b+c)+2} \times \frac{1}{\gon(a,b,c)}\]}\\
$\bullet$$\M(m,n,p)$ is the coefficient of $x^m y^n z^p$ in the expansion of $(x+y+z)^{m+n+p}$,  and $\M(m,n,p,1)=(m+n+p+1)\M(m,n,p)$,
therefore $gon(a,b,c)$ is equal to $(\sigma +1)$ times the coefficient of $x^{\sigma -a} y^{\sigma-b} z^{\sigma -c}$ in the expansion of $(x+y+z)^\sigma$ where $\sigma$ is the semi-perimeter of the triangle $(a,b,c)$.

\subsubsection{$SU(2)$ O-blades and  honeycombs} 
\label{Oblades}

\begin{figure}[htb]
\begin{center}
\includegraphics[width=0.2\textwidth]{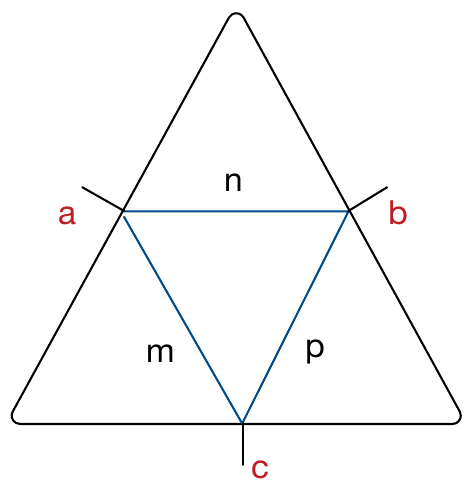}
$\qquad$
\includegraphics[width=0.2\textwidth]{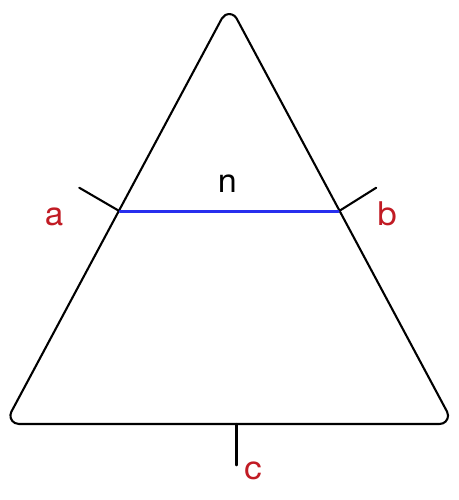}
\caption{Left: A generic O-blade for $SU(2)$. Right: A degenerate case $a=b$, $c=0$.} 
\label{oblade_su2}
\end{center}
\end{figure}

In sec.~\ref{admissibletriple} we called external variables the integers $(a,b,c)$ labelling an admissible triplet, and internal variables the integers $(m,n,p)$ that are such that $a=m+n, b=n+p, c=p+m$. 
If $a,b,c$ are thought as labels for highest weights of $\SU(2)$ irreps, we know that the concept of admissibility can be interpreted as coming from the requirement that  $c$ appears in the decomposition of $a \otimes b$ as a sum of irreps.
Replacing $SU(2)$ by $SU(N)$, $N \geq 2$, leads to analogous considerations but for the fact that multiplicities appearing in the decompositions of the tensor products of irreps can be larger than $1$; this difficulty triggered the invention of several types of pictographs 
(Berenstein-Zelevinsky diagrams \cite{BZ}, Ocneanu O-blades  \cite{AOblades}, Knutson-Tao honeycombs  \cite{KT1}) that are such that the number of distinct pictographs, for a given admissible triplet of highest weights $(a,b,c)$, is precisely equal to the associated multiplicity.
 This is not the right place to give the definitions and properties of these combinatorial tools but we want to mention that the admissible triangles considered previously can be understood as particular cases of such pictographs when the chosen Lie group is $\SU(2)$.
 For illustration, we display a generic $\SU(2)$ O-blade in fig.~\ref{oblade_su2}; notice that  $(a,b,c)$ indeed labels its external edges and $(m,n,p)$ its internal ones. 
  For $\SU(N)$ O-blades with $N>2$,  the number of internal edges  is larger than the number of components of the three chosen highest weights, but for $N=2$, they are equal.

  {   In the $SU(2)$ case the above picture also provides a combinatorical interpretation for the function $\gon(a,b,c)$.
  Let us replace a single internal edge labelled $n$ by $n$ parallel strands, or wires, (with an analogous replacement for $m$ and $p$), this leads to a picture where we have distributed $\sigma=m+n+p$ distinct wires $w_j$ among three directions $A,B,C$, with $m$ objects along the first direction, $n$ objects along the second, and $p$ objects along the third. The number of such partitions is the multinomial coefficient $M(m,n,p)$ and each choice can be specified by an anagram of the word $A(m\, {\text times}) B(n\, {\text times}) C(p\, {\text times})$ or by a triple $A(s_1), B(s_2), C(s_3)$ where the $s_i$ are increasing sequences of integers of lengths $m,n,p$ building a set partition of $\{1,2,\ldots, \sigma\}$, for instance $A(2), B(1,4), C(3,5,6)$.  The extra factor $(m+n+p+1)$ leading from $M(m,n,p)$ to $M(m,n,p,1)$, i.e. to $\gon(a,b,c)$ comes from a marking of the wires (or the choice of an orientation) with the constraint that, for each partition, if the wire $w_j$ is marked,  all wires  $w_j$ with $i<j$ should be marked as well.
  Parallel wires can be interpreted as tensor products of copies of the fundamental representation of $SU(2)$ giving rise, after appropriate symmetrization, to arbitrary irreducible ones; such graphical interpretations are among the foundational building blocks of topological quantum field theory (TQFT) and recoupling theory, see also our short discussion of Hom-spaces and intertwiners in \ref{Hom-spaces},  but here is not the right place to discuss those topics further.}  
 
\subsubsection{A topological identity of the $\gon$ function: the double triangle identity}
\label{sec:doubletriangleidentity}
{  The following proposition will not come as a surprise for people working in recoupling theory but we could not find a reference where it is explicitly stated.
The proof of this duality property is obtained after performing elementary changes of variables in the finite sums involved in  eq.~\ref{dualityidentity} or \ref{dualityidentityU}, below, and by using the definition of $\gon(a,b,c)$ as a multinomial, 
but one has to consider many cases that depend upon the relative values of $a$, $b$, $c$, $d$, and of their pair-wise sums or differences. 
The proof, that is left to the reader, is therefore elementary but rather lengthly and cumbersome.} 

\begin{figure}[htb]
\begin{center}
\includegraphics[width=0.5\textwidth]{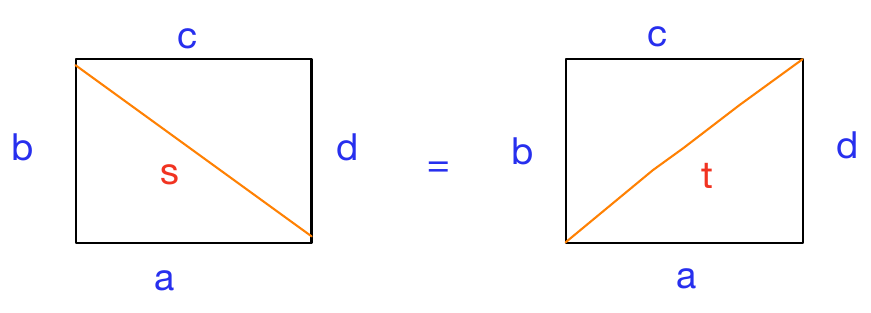}
\caption{The duality equation} 
\label{diamonds}
\end{center}
\end{figure}

Consider a quadrilateral $(a,b,c,d)$, it can be thought as the union of two triangles $(a,b,s)$ and $(c,d,s)$ glued along the diagonal $u$, or as the union of two triangles $(a,d,t)$ and $(b,c,t)$ glued along the other diagonal $t$ (fig.~\ref{diamonds}).
We have:
{\proposition{}
\begin{equation} \label{dualityidentity}
\sum_{s\in S} \gon(a,b,s) \frac{1}{s+1} \gon(c,d,s) = \sum_{t \in T} \gon(a,d,t) \frac{1}{t+1} \gon(b,c,t)  \end{equation}
In the first sum, $s$ runs over the set $S$ of integers such that both $(a,b,s)$ and $(c,d,s)$ are admissible triplets; in second sum, $t$ runs over the set $T$ of integers making admissible both $(a,d,t)$ and $(b,c,t)$.
The above sums are also equal to \begin{equation}\sum_{u \in U} \gon(a,c,u) \frac{1}{u+1} \gon(b,d,u)\label{dualityidentityU}\end{equation} where $u$ runs over the set $U$ of integers making admissible both  $(a,c,u)$ and $(b,d,u)$.\\
{  Quadrilateral function:} the left or right hand side of eq.~\ref{dualityidentity}, or the expression \ref{dualityidentityU}, defines the number $\gon(a,b,c,d)$ which is an integer for each quadruplet.}\\
About integrality: We already know that  $\gon(a,b,c)$ is divisible by $(c+1)$,  therefore each term of the sums (\ref{dualityidentity}) or (\ref{dualityidentityU}) is also an integer, and, consequently, their sum $\gon(a,b,c,d)$ as well.

{  To illustrate the above proposition let us take $(a,b,c,d) = (3,4,6,11)$, therefore $S=\{5, 7\}$, $T=\{8,10\}$, $U=\{7,9\}$; the sum over $S$ reads $388080 + 2522520$, the sum over $T$ reads $1108800+1801800$, the sum over $U$ reads $748440+2162160$, and the three sums are indeed equal.}

In terms of $SU(2)$ representation theory, using integers labelling (highest weights of) irreducible representations, $S$ denotes the intersection of the set of irreps that appear {  in the decomposition into irreps} of the tensor products $a\otimes b$ and $c\otimes d$,  $T$ denotes the intersection of the set of irreps that appear in $a\otimes d$ and $b\otimes c$, and $U$ denotes the intersection of the set of irreps that appear in $a\otimes c$ and $b\otimes d$.
Using the notation (\ref{bigotimesset}) we have explicitly  $S=\bigotimes[a,b] \cap  \bigotimes[c,d]$, $T=\bigotimes[a,d] \cap \bigotimes[b,c]  $, and $U=\bigotimes[a,c] \cap \bigotimes[b,d]$.

A quadruplet of non-negative integers (labelling irreps) is called admissible when the tensor product $a\otimes b\otimes c \otimes d$ contains the trivial representation {  ---we shall say more about this notion in sec.~\ref{generalpolygon}. 
If the quadruplet is not admissible, $\gon(a,b,c,d)$ vanishes.}

{   Intuitively, one either couples the irreps $a$ and $b$ to give the possible $s$'s that, themselves, couple to the irreps $c$ and $d$, or one first couples $a$ and $d$, to give the possible $t$'s that couple to $b$ and $c$.
This ``coupling'' can be interpreted in terms of composition of equivariant homomorphism (intertwiners) which are briefly discussed in section \ref{Hom-spaces}.} 
Using particle physicists parlance, one can say that $\gon(a,b,c,d)$  is invariant upon interchange of the $s$, $t$, and $u$ channels (the symbols $s, t, u$ refer to the Mandelstam variables introduced in the 50's to study duality properties in particle diffusion processes \cite{Mandelstam}).

{ 
A combinatorial interpretation of the above  quadrilateral function can also be given, along the same lines as the ones discussed for $\gon(a,b,c)$ itself in sec.~\ref{Oblades}, by drawing a square with either one diagonal or the other, 
by replacing edges and  the chosen diagonal by wires relating the four corners,
and interpreting each contribution to the above sums as a counting of the possible distributions of wire labels (obeying a marking constraint similar to the one discussed in \ref{Oblades})  among five directions: the four sides of the quadrilateral and the  chosen diagonal.

Some readers will have noticed that, up to a possible replacement of $\gon$ by the theta net of sec.~\ref{KauffmanTheta}, a single contribution to the above sums, like the term $\gon(a,b,s) \, \gon(s,c,d)/(s+1)$,  is ubiquitous in recoupling theory and in the theory of spin networks where it appears as a normalization coefficient for one of the ``double triangles''  that make a tetrahedron (see sec.~\ref{classicaladmissibletetrahedra}).}

\subsubsection{Definition of the function $\gon$ for arbitrary admissible polygons and multisets}
\label{generalpolygon}

{  
It is convenient in the present discussion to consider multisets  (in a multiset, order does not matter and repetitions are allowed).
We  denote multisets either by square brackets or by braces, like sets, while adding an index to elements to denote their multiplicity (which can be absent it is equal to $1$), for instance $[2,2,2,3,3]=\{2_3,3_2\}$.
Consider now a multiset $[a_i]$ of non-negative integers such that $\sum a_i$ is even and $2\, {\max}(a_i) \leq \sum a_i$.  Such a multiset will be called admissible.
We define recursively the function $\gon$ on arbitrary admissible multisets as follows:
{\definition
Let $[a_i]$ be an admissible multiset and $u$, $v$, two non-negative integers such that the multiset $[a_i,u,v]$ obtained by adjoining $u$ and $v$ to $[a_i]$ is also admissible. Then we set
\begin{equation}
\gon(a_i, u, v) = \sum_x \gon(a_i, x) \frac{1}{(x + 1)} \gon(x, u, v), 
\label{gononmultisets}
\end{equation} 
 where $x$ belongs to the intersection of the sets $\bigotimes[u,v]$ and $\{m, m+2,m+4,m+6\ldots,  \max(a_i)\}$, where $m = \max(2\, \max(a_i) -  \sum a_i,0)$.}
 
  This formula defining $\gon$ on the admissible multiset $[a_i, u, v]$ assumes that $[a_i]$  possesses at least two elements; the base case of the recursion is known from the already defined function $\gon$ on admissible triplets.
 One could remove the admissibility requirement from the above definition, indeed the value of $\gon$ on a non admissible multiset would automatically vanish because of the appearance of non admissible triplets on the right hand side of eq.~\ref{gononmultisets}.
 For this definition to make sense for a given admissible multiset, i.e.~for the obtained function to be a symmetric function of its arguments, one has to show that the result is independent of the method used to recursively construct this multiset in terms of smaller ones; 
 the proof (left to the reader) comes from the double triangle identity and from the symmetry property of the elementary $\gon$ function. 
 
 From a group theoretical point of view, the  condition of admissibility for a multiset $[a_i]$ can be interpreted, or defined, from the requirement that the tensor product  $\otimes a_i$ of the $SU(2)$ irreps defined by the $a_i$'s (thought of as highest weights) should contain the trivial representation.  The two sets whose intersection has to be taken to select the $x$'s that enter  eq.~\ref{gononmultisets} have the following interpretation: 
 the first set is the set of highest weights of those irreps of $SU(2)$ that appear in the decomposition of $u\otimes v$ into irreducible components; 
 the second is the set of (distinct) highest weights that appear in the decomposition of $\otimes a_i$ into irreps, equivalently it is the underlying set of the multiset $\bigotimes[a_i]$;
  it can be determined iteratively by using associativity of the representation ring.
 The obtained list of highest weights builds a multiset $\bigotimes[a_i]$ with usually non-trivial multiplicities since, for a product involving more than two irreps, the multiplicities can be bigger than $1$ but in the recursive definition (\ref{gononmultisets}) one does not take  multiplicities into account.
The factor $(x+1)$ in the denominator can be interpreted as the dimension of the representation labelled by the highest weight $x$.

The definition of $\gon$ given in eq.~\ref{gononmultisets} has a natural geometrical interpretation that can also be used as a definition.}
Consider an arbitrary polygon with (non-negative) integer sides, we define the function $\gon$ on this polygon as follows.
By inserting diagonals, the polygon can be triangulated in many ways. {  Choose some triangulation and label the diagonals $c$ in such a way that $c$ belongs to the set $\bigotimes[a,b]$ for each of the triangles $\{a,b,c\}$ that countain $c$ (at most $2$); this defines a labelling of the triangulation.}
 {  For a given labelling,} take the product of the triangular $\gon$ functions over all the 2-simplices of the triangulation, divided by a product of factors $(x+1)$ where the $x$'s  are the lengths of the diagonals (1-simplices) entering the chosen triangulation; 
 {  the last step is to sum this product over all possible labellings of the chosen triangulation. The duality property of the elementary $\gon$ function, interpreted geometrically as in fig.~\ref{diamonds}, ensures that the obtained result is independent of the chosen triangulation.
 The  polygon is called non-admissible if the obtained value is equal to $0$ (this can be traced back to the occurence of non-admissible triangles in the triangulation). 
 The obtained $\gon$ function is symmetric as a function of the edges of the polygon (proof: use this symmetry for the known elementary $\gon$ function defined on triangles).
 }

\begin{figure}[htb]
\begin{center}
\includegraphics[width=0.6\textwidth]{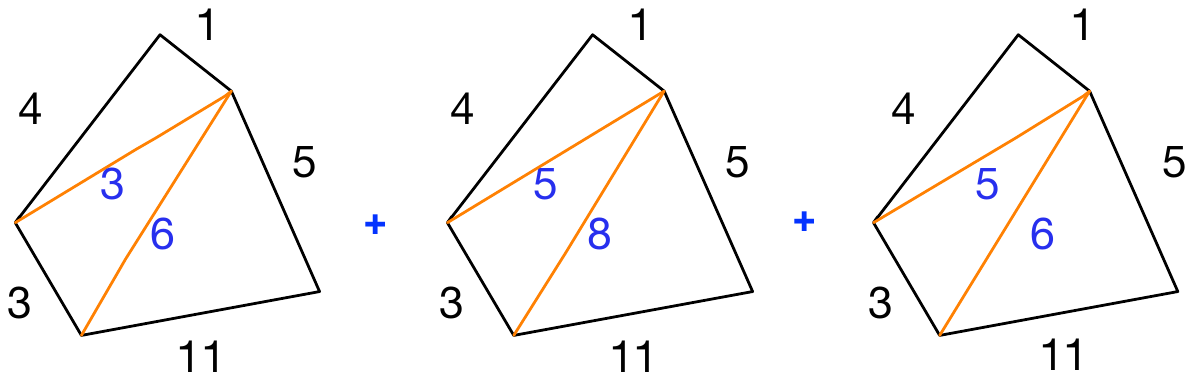} $=18295200$
\caption{Evaluation of $\gon$ on the $5-plet$ $(11,3,4,1,5)$ (pentagon)} 
\label{pentagone_triangles}
\end{center}
\end{figure}

\begin{figure}[htb]
\begin{center}
\includegraphics[width=0.4\textwidth]{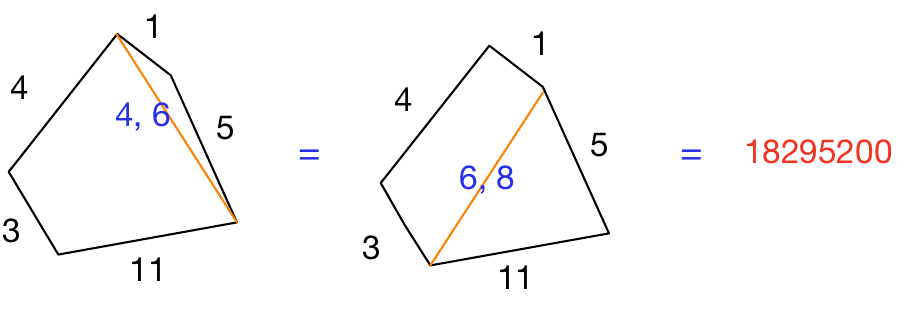} $=18295200$
\caption{Other evaluations of $\gon$ on $(11,3,4,1,5)$} 
\label{pentagone_quadri}
\end{center}
\end{figure}

A pentagonal example. Consider the multiset $[11, 3, 4, 1, 5]$.
It is indeed admissible and the value of $\gon$ can be obtained by introducing two diagonals like in fig~\ref{pentagone_triangles} and summing over the three triangles.
Using {\scriptsize $\otimes[1,4]=\{3,5\}$,   $\otimes[3,5]=\{2,4,6,8\}$,  $\otimes[5,11]=\{6,8,10,12,14,16\}$}, one finds\\
  {\scriptsize $\gon(11, 3, 4, 1, 5)=$}
$\frac{\gon(6, 5, 11) \gon(6, 3, 3) \gon(3, 4, 1)}{(6 + 1) (3 + 1)} +  \frac{\gon(1, 4, 5)}{(5+1)} (\frac{\gon(3, 5, 8) \gon(8, 11, 5)}{(8+1)} + \frac{\gon(3, 5, 6) \gon(6, 11, 5)}{(6+1)})$ {\scriptsize $=18295200$}.\\
One could instead use eq.~\ref{gononmultisets}, first calculating $\gon$ on the quadrilateral of fig~\ref{pentagone_quadri} (left) and completing the picture by adding one single triangle with two possible values ($4$ and $6$) 
for the diagonal, indeed {\scriptsize $\bigotimes[5,1]=\{4,6\}$} and {\scriptsize $\bigotimes[4,11,3]=\{4_1, 6_2, 8_3, 10_4, 12_4, 14_3, 16_2, 18_1\}$}.
The same value of $\gon$ would be obtained as
$\frac{\gon(11, 3, 4, 4) \gon(4, 1, 5)}{(4 + 1)} +  \frac{\gon(11, 3, 4, 6) \gon(6, 1, 5)} {(6 + 1)}$.
We could also choose another diagonal, like in fig~\ref{pentagone_quadri} (right), and calculate
$\frac{\gon(3, 4, 1, 6) \gon(6, 5, 11)}{(6 + 1)}+ \frac{\gon(3, 4, 1, 8) \gon(8, 5, 11) }{(8 + 1)}$ since the possible values of this diagonal are $6$ and $8$, indeed, {\scriptsize $\bigotimes[11,5]=\{{6, 8, 10, 12, 14, 16}\}$} and {\scriptsize $\bigotimes[3,4,1]=\{0_1, 2_2, 4_2, 6_2, 8_1\}$}.
The displayed polygons may be misleading since most triangles of this  triangulation are degenerate -- the figures are not metrically correct.
One should also remember that $\gon$ is symmetric, the order of arguments (sides of the n-gon) is therefore irrelevant.

\subsubsection{The $\ThetaGraph_K$ function} 
\label{KauffmanTheta}

In the Temperley-Lieb recoupling theory, in topological field theory, or in the theory of spin networks, one introduces the following function $\ThetaGraph_K$,  \cite{KauffmanLins} : 
\begin{equation}
\ThetaGraph_K (a,b,c)  = (-1)^{m+n+p}    \frac{(m+n+p+1)! \,  m! n! p! }{(m+n)! (m+p)! (n+p)!} \label{classicalthetadef} \end{equation} 
 {  This function is usually denoted $\Theta$ and often called ``the theta net'', but there are several possible evaluation procedures in the theory of spin networks, this is why we  added the suffix $K$ to the $\Theta$ symbol (see our discussion in section \ref{sec:classicalspinnetworks}).
It is also sometimes called the trihedron coefficient \cite{MasbaumVogel}.}
Its relation with the $\gon$ function is immediate. One has: 
\begin{equation}
{\ThetaGraph}_K (a,b,c)   = (-1)^{m+n+p}   \frac{ \M(m,n,p,1) }{\M(m,n)  \M(n,p)  \M(p,m)}  = (-1)^{\frac{a+b+c}{2}}   \frac{\gon(a,b,c)}{\binom{a}{\frac{1}{2} (a+b-c)}  \binom{b}{\frac{1}{2} (b+c-a)}   \binom{c}{\frac{1}{2} (c+a-b)} } 
\label{gontotheta}
\end{equation}
$M$ with two arguments denotes the binomial coefficient, for instance $\M(m,n) = \binom{m+n}{n} = \binom{a}{\frac{1}{2} (a+b-c)}$.

Notice that $\text{gon}(a,b,c)$ is a positive integer whereas ${\ThetaGraph}_K (a,b,c) $ is a rational number that can have both signs, it is usually not an integer.

In the degenerate case $a=b$, $c=0$, one has $m=p=0$, $n=a=b$, so ${\ThetaGraph}_K (a,a,0) = (-1)^a (a+1)$, which, up to sign, is the dimension of the $\SU(2)$ irrep with highest weight $a$ {  (this signed quantity is often denoted $\Delta(a)$ in the literature).}

\subsubsection{$\gon(a,a,b,b)$ and the hypergeometric function}
{  Let us find a closed expression for $\gon(a,a,b,b)$.
The intersection of  $\bigotimes[a,a]=\{0,2,\ldots,2a\}$ and $\bigotimes[b,b]=\{0,2,\ldots,2b\}$ is $\{0,2,\ldots,2\min(a,b)\}$. 
From the recursive definition of the $\gon$ function, we have
$\gon(a,a,b,b) =  \sum_{t=0, \; t\,\text{even}}^{2 \min (a,b)}  \gon(a,a,t) \, \frac{1}{t+1}\, \gon(t,b,b)$, where $\gon(a,a,t)$ and $\gon(t,b,b)$ are already known, therefore}
\begin{equation}
\gon(a,a,b,b) =  \sum_{t=0, \; t\,\text{even}}^{2 \min (a,b)} \,
\frac{\left(\frac{1}{2} (2 a+t)+1\right)!}{\left(\frac{t}{2}!\right)^2 \left(\frac{1}{2} (2 a-t)\right)!}\,
\frac{1}{t+1}\,
\frac{\left(\frac{1}{2} (2 b+t)+1\right)!}{\left(\frac{t}{2}!\right)^2 \left(\frac{1}{2} (2 b-t)\right)!}
\label{gonabab:sum}
  \end{equation}
{  By using the series expansion of the hypergeometric function $_pF_q$ one can rewrite the above finite sum as follows:}
\begin{equation}\gon(a,a,b,b) = (a+1) (b+1) \; _5F_4\left(1-(a+1),1-(b+1),1+(a+1),1+(b+1),\frac{1}{2};1,1,1,\frac{3}{2};1\right)  \label{gonaabbhyperg} \end{equation}
 In the particular case with $a=b$, and setting $j=t/2$ one obtains 
    \begin{equation}
    \gon(a,a,a,a)= \sum_{j=0}^a  \frac{\Gamma \left(a+j+2\right)^2}{(2j+1) \Gamma \left(j+1\right)^4 \Gamma \left(a-j+1\right)^2}
    \label{gonaaaahyperg}
     \end{equation}
The sequence $a\in \N \mapsto  \gon(a,a,a,a) \in \N$ starting from $a=0$, namely
$1, 16, 381, 10496, 307505$, $9316560, 288307285, 9052917760, 287307428985, 9192433560080, \ldots$
 can be recognized as the sequence  OEIS A189766.
 
\subsubsection{The function $\gon$ and Hilbert matrices}
\label{sec:Hilbert}

Remember that the Hilbert matrix $H$ of order $n$ is defined as the $n\times n$ matrix $H(n)$ with matrix elements $H(n)(i,j) = 1/(i+j-1)$. \\ For instance {\scriptsize 
\[
H(3)= 
\left(
\begin{array}{ccc}
 1 & \frac{1}{2} & \frac{1}{3} \\
 \frac{1}{2} & \frac{1}{3} & \frac{1}{4} \\
 \frac{1}{3} & \frac{1}{4} & \frac{1}{5} \\
\end{array}
\right)
\]}

{\proposition{}The value $gon(n,n,n,n)$ is equal to  the trace of the inverse of the $(n+1)$-th order Hilbert matrix.}

\smallskip
Example : 
{\scriptsize
\[
{H(3)}^{-1}=
\left(
\begin{array}{ccc}
 9 & -36 & 30 \\
 -36 & 192 & -180 \\
 30 & -180 & 180 \\
\end{array}
\right),
\qquad
\tr({H(3)}^{-1})=381,
\qquad
\gon(2,2,2,2)=381.
\]
}
{  Entries of the inverse of the Hilbert matrix can be expressed in closed form using binomial coefficients (see for instance \cite{HilbertMatrix_wiki}), summing over diagonal elements leads to eq.~\ref{gonaaaahyperg}, hence the result. 
Another way to prove this result is to rely on the OEIS data basis since  one of the expressions of  $\tr(H(n+1)^{-1})$ for the generic term of the sequence  A189766 of  \cite{OEIS}, written in terms of an hypergeometric function,  coincides  with (\ref{gonaabbhyperg}) when $a=b$}.
Admittedly, a more conceptual proof should be looked after.

\smallskip
We now define the {\sl shifted Hilbert matrix of order $n$, and shift $s\in \N$}, as the  $n\times n$  matrix $H(n,s)$ with matrix elements
$H(n,s)(i,j)=1/(i + j - 1 + s).$ \\ For instance {\scriptsize  
\[
H(5,3)=
\left(
\begin{array}{ccccc}
 \frac{1}{4} & \frac{1}{5} & \frac{1}{6} & \frac{1}{7} & \frac{1}{8} \\
 \frac{1}{5} & \frac{1}{6} & \frac{1}{7} & \frac{1}{8} & \frac{1}{9} \\
 \frac{1}{6} & \frac{1}{7} & \frac{1}{8} & \frac{1}{9} & \frac{1}{10} \\
 \frac{1}{7} & \frac{1}{8} & \frac{1}{9} & \frac{1}{10} & \frac{1}{11} \\
 \frac{1}{8} & \frac{1}{9} & \frac{1}{10} & \frac{1}{11} & \frac{1}{12} \\
\end{array}
\right)
\]
}
One has $H(n,0)=H(n)$.
One may consider $H(n,s)$ as a $n\times n$ submatrix of the infinite Hilbert matrix, with the diagonal element $1/(s+1)$ as its upper left corner.

\smallskip
Notice that {\scriptsize 
$
{H(5,3)}^{-1}=
\left(
\begin{array}{ccccc}
 19600 & -141120 & 352800 & -369600 & 138600 \\
 -141120 & 1058400 & -2721600 & 2910600 & -1108800 \\
 352800 & -2721600 & 7144200 & -7761600 & 2993760 \\
 -369600 & 2910600 & -7761600 & 8537760 & -3326400 \\
 138600 & -1108800 & 2993760 & -3326400 & 1306800 \\
\end{array}
\right)
$
}

{  
Summing over all entries of the diagonal of the inverse of the shifted Hilbert matrix (they can again be deduced from  \cite{HilbertMatrix_wiki}) leads to eq.~\ref{gonabab:sum}.
Summing over appropriate rows (see below) leads to a multinomial coefficient that one recognizes as the value of $\gon(a,b,c)$. In this way one obtains the next two propositions:
}

{\proposition{} For all non-negative integers $a,b$, one has 
\[\gon(a, b, a, b) = \tr({H(\min(a,b) + 1, |a-b|)}^{-1})\]}
Example: $\gon(4, 7, 4, 7)= tr({H(5,3)^{-1}})=18066760.$

\bigskip
The value $\gon(a,b,c)$ itself can be obtained from the inverse of a shifted Hilbert matrix by summing the matrix elements of an appropriate row:
{\proposition{}
Without loss of generality, let us assume that the admissible triplet $(a,b,c)$ is such that $a \leq b \leq c$, then
one has 
\[\gon(a, b, c)= \bigg\lvert \sum_{} H(a+ 1, b-a)^{-1}[[(c + a - b)/2 + 1]] \bigg\rvert \]}
where $M[[j]]$ denotes the row $j$ of a matrix $M$, and $\sum M[[j]]$ the sum over the corresponding matrix elements.
Warning: we know that $\gon$ is symmetric in its arguments, but the right hand side of the previous equality is not, hence the importance of ordering the arguments $a,b,c$ as indicated.

\smallskip
Example:  $\gon(4,7,9)=9240$, to be compared with\\
 $\sum_{} H(5, 3)^{-1}[[4]]=\sum\{-369600, 2910600, -7761600, 8537760, -3326400\}=-9240$.

\subsubsection{Other properties of the function $\gon$} 

\smallskip
{\bf From the function $\gon$ to special Clebsch-Gordan  and Wigner $3j$ symbols.}
{   The relation expressing Clebsch-Gordan coefficients in terms of $3j$ symbols (and conversely) is recalled  in the appendix, eq.~\ref{3jtoCG}.
 We write it  here using  in-line Mathematica notations.}
{\footnotesize
   \begin{equation}
\text{ClebschGordan}(\{{j_1},{m_1}\},\{{j_2},{m_2}\},\{j,m\})=\sqrt{2 j+1} (-1)^{{j_1}-{j_2}+m}
   \text{ThreeJSymbol}(\{{j_1},{m_1}\},\{{j_2},{m_2}\},\{j,-m\}) \end{equation}}
The arguments $j_1, j_2$, and $j$, in this formula, are spin variables (half-integers).
Remember that the arguments of the $\gon$ functions, called $a,b,c,\ldots$ in the previous sections, are twice the spin variables.
The spin $j$ is allowed to take values from  $|j_1-j_2|$ to $j_1+j_2$ and the (half-integer) arguments $m_i$ run from $-j_i$ to $+j_i$, one has also  $m=m_1+m_2$, otherwise the Clebsch-Gordan coefficient vanishes.
If both $j_1$ and $j_2$ are integers (the so-called orbital case), $j$ has to be an integer as well, and choosing arguments  $m_1=m_2=m=0$ is allowed; one can therefore consider the particular values 
  {\footnotesize 
   \begin{equation}
    \label{special_Clebsch_def}
   \text{ClebschGordan}(\{{j_1},{0}\},\{{j_2},{0}\},\{j,0\})=\sqrt{2 j+1} \, (-1)^{{j_1}-{j_2}} \,   \text{ThreeJSymbol}(\{{j_1},{0}\},\{{j_2},{0}\},\{j,0\})
  \end{equation}}
 {  An explicit expression for (\ref{special_Clebsch_def}) was obtained long ago,
  see formula 3.194 of the book \cite{BiedenharnLouck}, vol VIII,} the authors attribute this expression to Racah --- notice that they call ``Wigner coefficients" what everybody  nowadays would call ``Clebsch-Gordan coefficients".
{  Using  eqs.~(\ref{gonexpression}), one can then rewrite their expression in terms of the $\gon$ function and therefore express the above particular Clebsch-Gordan coefficients as follows}:
   {\footnotesize
 \begin{equation}
 \label{special_Clebsch}
 \text{ClebschGordan}(\{j_1,0\},\{j_2,0\},\{j,0\})=    \cos \left( (j_1+j_2-j)\frac{\pi}{2} \right)   \;   \;  \sqrt{2 j+1}  \;  \frac{1}{\left(\frac{1}{2} (j+j_1+j_2)+1\right)}  \;   \frac {\gon(j_1,j_2,j)} {\sqrt{\gon(2 j_1,2 j_2,2 j)}}
  \end{equation}}
Notice that the $\cos$ term in (\ref{special_Clebsch}) is either $0$, or a sign ($\pm 1$), and that
 these special Clebsch-Gordan coefficients vanish
 if $(j_1,j_2,j)$ is not an admissible triple, in particular if $j_1+j_2+j$ is odd.
From a computational point of view, their evaluation in Mathematica using the function $\gon$  (as defined  in section \ref{gon:definition}) is about ten times faster than the one using the built-in \text{ClebschGordan} function.

Notice that since we know how to write the $\gon$ function (with three arguments) in terms of sums of coefficients of an inverse and shifted Hilbert matrix, 
one can also relate the special Clebsch-Gordan coefficients  (\ref{special_Clebsch_def})  to Hilbert matrices by using the formulae of section \ref{sec:Hilbert}.

\smallskip
{\bf An expression of $\gon$ using a Dyson - MacDonald identity.} 
 {  This identity states that the  Laurent polynomial   $\prod_{1\leq i \neq j \leq k} (1 - t_i/t_j)^{s_i}$ has constant term equal to the multinomial coefficient $\text{M}(s_1,s_2,\ldots, s_k)$.}
 It was first conjectured by Dyson \cite{Dyson} and later proved by Wilson \cite{WilsonKG}, Gunson \cite{Gunson}, Good \cite{Good}  (the conjecture was then generalized by MacDonald \cite{MacDonald}, with a general proof given by Cherednik \cite{Cherednik}).
 
With $m,n,p$ defined in terms of $a,b,c$ as in \ref{admissibletriple}, one can use this identity with $k=4$ to express the function $\gon(a,b,c)$,  written  as  the multinomial coefficient with four parts $M(m,n,p,1)$, as the constant term of a Laurent polynomial; 
{  one can also set $t_1 = u v w\, t_4, t_2 =  v w\, t_4, t_3 = w\, t_4$ and trade the four variables $t_i$ for the three variables $u,v,w$.  Therefore we have:}\\
{\proposition{} :
 $\gon(a,b,c)$  can be obtained as the constant term of (\ref{gonDyson1}) or (\ref{gonDyson2})}
 {\footnotesize
 \begin{equation}
\left(1-\frac{1}{w}\right) (1-u)^{{m}} \left(1-\frac{1}{u}\right)^{{n}} (1-v)^{{n}} \left(1-\frac{1}{v}\right)^{{p}} (1-w)^{{p}}
   \left(1-\frac{1}{v w}\right) (1-u v)^{{m}} (1-v w)^{{n}} \left(1-\frac{1}{u v}\right)^{{p}}
   \left(1-\frac{1}{u v w}\right) (1-u v
   w)^{{m}} 
\label{gonDyson1}
 \end{equation}
   }
{  One could also write  $\gon(a,b,c)$ as $(m+n+p+1)$ times  the multinomial coefficient with three parts $M(m,n,p)$ and use the Dyson-MacDonald identity with $k=3$ to write the latter as the constant term of another multivariate Laurent polynomial. 
One can then set $t_1 = u v  \, t_3, t_2 =  v\,  t_3$ and trade the three variables $t_i$ for the two variables $u,v$.  With that choice, $\gon(a,b,c)$  is recovered as the constant term of 
 {\footnotesize
 \begin{equation}
 (m+n+p+1) \left(1-\frac{1}{u}\right)^{n} \left(1-\frac{1}{v}\right)^{p} \left(1-\frac{1}{u v}\right)^{p} (1-u)^{m}  (1-v)^{n} (1-u v)^{m}\
\label{gonDyson2}
 \end{equation}}
Admittedly these expressions are a bit heavy and more complicated than the explicit result itself; 
one hope was to use them to obtain similar expressions for the multivariable generalizations of $\gon$, but this is not known to the author.}

\smallskip
{\bf Asymptotics of $\gon$.} 
Using the Stirling formula for the factorial leads to the following
 {\proposition{}: When $k$ goes to infinity, we have
   \begin{equation}\gon(k a,k b, k c) \; \sim_{k \rightarrow \infty} \; \frac{\sigma^2}{2 \pi  A}   \left(\frac{\sigma^\sigma}{m^m n^n p^p}\right)^k   \end{equation}
where $A = \sqrt{\sigma \, m \, n \, p}$ is the area (Heron formula),  $\sigma$ is the semi-perimeter of the triangle $(a,b,c)$, and $m,n,p$ are the usual variables $m=\sigma - b, n = \sigma-c, p = \sigma -a$.}
The next term of the above expansion is of order $O(1/k)$.

One can work out the asymptotic expansion of $\ThetaGraph_K$  as well, and find
\[ \ThetaGraph_K(k a,k b, k c) \; \sim_{k \rightarrow \infty} \;   (-1)^{k\, \sigma} \frac{k^{3/2} \sigma A \sqrt{2 \pi }}{\sqrt{(m+n) (m+p) (n+p)}}  \, \left(\frac{\sigma^\sigma m^m n^n p^p}{(m+n)^{m+n} (n+p)^{n+p} (p+m)^{p+m}}\right)^k
\]

\subsection{The classical $\tet$ function}

\subsubsection{Admissible tetrahedra}
\label{classicaladmissibletetrahedra}

\begin{figure}[htb]
\begin{center}
\includegraphics[width=0.3\textwidth]{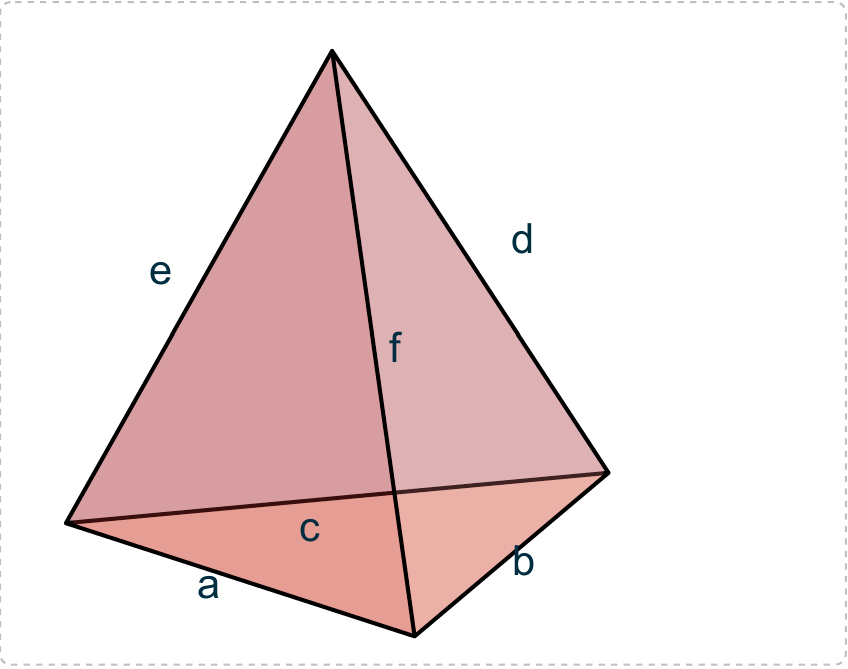}
\caption{A labelled tetrahedron (external variables are associated with its edges).} 
\label{tetrahedronT}
\end{center}
\end{figure}

Consider $((a,b,c),(d,e,f))$, a pair of triplets of non-negative integers. We shall say that this  defines an admissible tetrahedron $T$ if the following constraints are obeyed:
the first triplet $(a,b,c)$ is an admissible triple, as defined in section \ref{admissibletriple}, and the triplets $(b,d,f)$, $(a,e,f)$, $(c,d,e)$ are also  admissible. 
Warning: the triplet $(d,e,f)$ that appears as the second argument of $T$ is usually not admissible.  We shall often write this  tetrahedron as \[T=\big(\begin{smallmatrix} a & b & c\\ d & e & f \end{smallmatrix}\big).\]
Such a pair of triplets (the first being admissible) can be displayed as a tetrahedron $T$, as follows:  
the components of the first argument, \ie the triplet $(a,b,c)$, refer to edges defining one of the triangular faces of $T$, and the components of the second argument $(d,e,f)$, in this order, denote respectively the edges of $T$ that are skew (\ie opposite) with respect to $a,b,c$,
see fig.~\ref{tetrahedronT}.
Obviously, the same tetrahedron can be obtained by choosing, as a first argument, any of its four triangular faces, and as a second argument, the triplet of edges that are skew with respect to the three edges of the chosen face. 
The above notation for a tetrahedron being not unique we think of $T$ as an equivalence class of such pairs of triplets under the action of the tetrahedral group.

The triangular conditions on faces ensure that the four faces of $T$ can indeed be realized as actual euclidean triangles. 
Warning: these conditions are not sufficient to imply that an admissible tetrahedron, in the above sense,  can be realized as an euclidean tetrahedron of $\R^3$, indeed, the Cayley-Menger determinant defined by the edges may not be positive.
Notice also that our ``edges'' may vanish (degenerated cases).  For all these reasons one should maybe write ``possibly degenerated admissible integer-sided non-congruent facial sextuples'' rather than ``admissible tetrahedra" but we shall nevertheless stick to the shorter terminology.

Removing one column in the symbol  $\big(\begin{smallmatrix} a & b & c\\ d & e & f \end{smallmatrix}\big)$ of $T$  (there are three possibilities) gives a labelling for the three closed skew quadrilaterals, for instance 
 $\big(\begin{smallmatrix} a & b \\ d & e\end{smallmatrix}\big)$ refers to the quadrilateral $(a,b,d,e)$. {  This gives rise to the so-called recoupling coefficients, see sec.~\ref{otherpropertiesoftet}}.

\subsubsection{The {  integer valued} $\tet$ function (the tetrahedral function, also denoted ${\tetrahedron}_\Z$)}  
\label{hed_function_def}
We shall first define a function $\tet$  on the set of admissible tetrahedra $T$ (if the argument is not admissible, the value of $\tet$ is set to $0$).

For the four triangular faces of $T$ we introduce the semi-perimeter variables 
    \begin{equation}\sigma(1)=\frac{1}{2} (a+b+c); \sigma(2)=\frac{1}{2} (b+d+f); \sigma(3)=\frac{1}{2}
   (a+e+f); \sigma(4)=\frac{1}{2} (c+d+e).    \end{equation}

   For the three closed (and skew) quadrilaterals  defined from the edges of $T$ we also introduce the semi-perimeter variables 
    \begin{equation}  \tau(1)=\frac{1}{2} (a+b+d+e);\tau(2)=\frac{1}{2} (a+f+d+c);\tau(3)=\frac{1}{2} (e+f+b+c). \end{equation}
   
    Moreover we set
   \[m_\sigma=\max (\sigma(1),\sigma(2),\sigma(3),\sigma(4)); \quad m_\tau=\min(\tau(1),\tau(2),\tau(3)).\]
 
    Then we define:
    {\definition
    \begin{equation}
       \tet(T)=
   \left(
   \sum _{s=m_\sigma}^{m_\tau} 
   \frac{(-1)^s  (s+1)!}
   {\left(\prod _{i=1}^4 (s-\sigma(i))!\right)  
   \left(\prod _{u=1}^3 (\tau(u)-s)! \right)}
   \right)
    \label{hed_definition}
        \end{equation}}
        
    {  The above expression, taken here as a definition of $\tet$, is simply related to the known value of the so-called tetrahedral net; this will be discussed in the next subsection.}  
         From the definition of the four $\sigma(i)$ variables (triangles) and of the three $\tau(u)$ variables (quadrilaterals), one finds $\sum_i \sigma(i) = \sum_u \tau(u) = a+b+c+d+e+f$, therefore $\sum_i (s-\sigma(i)) + \sum_j (\tau(j) - s) +1 = s+1$.
The previous expression can therefore be rewritten as a signed sum of multinomial coefficients: 
     \begin{equation}
       \tet(T)=
   \sum _{s=m_\sigma}^{m_\tau}  (-1)^s \, M(s-\sigma(1),s-\sigma(2),s-\sigma(3),s-\sigma(4),\tau(1)-s,\tau(2)-s,\tau(3)-s,1)
       \label{hed_definition_bis}
        \end{equation}   
 From the right hand side of eq.~\ref{hed_definition}, or of   eq.~\ref{hed_definition_bis} it is clear that the values obtained for the double triplets   $((a,b,c),(d,e,f))$,  $((b,d,f),(e,a,c))$, $((a,f,e),(d,c,b))$, $((c,d,e),(f,a,b))$, are the same, so that $\tet$ is indeed a function of $T$.
Since we can rewrite this expression as a sum of multinomials, it is clear that $\tet$ is integer-valued.

Examples:\\
 $\tet((8, 20, 24), (15, 13, 17))=332385335268386400$,\\
$\tet((14, 41, 33) (50, 23, 21))= -671777611858249170324639542553600$.

\bigskip

\begin{figure}[htb]
\begin{center}
\includegraphics[width=0.3\textwidth]{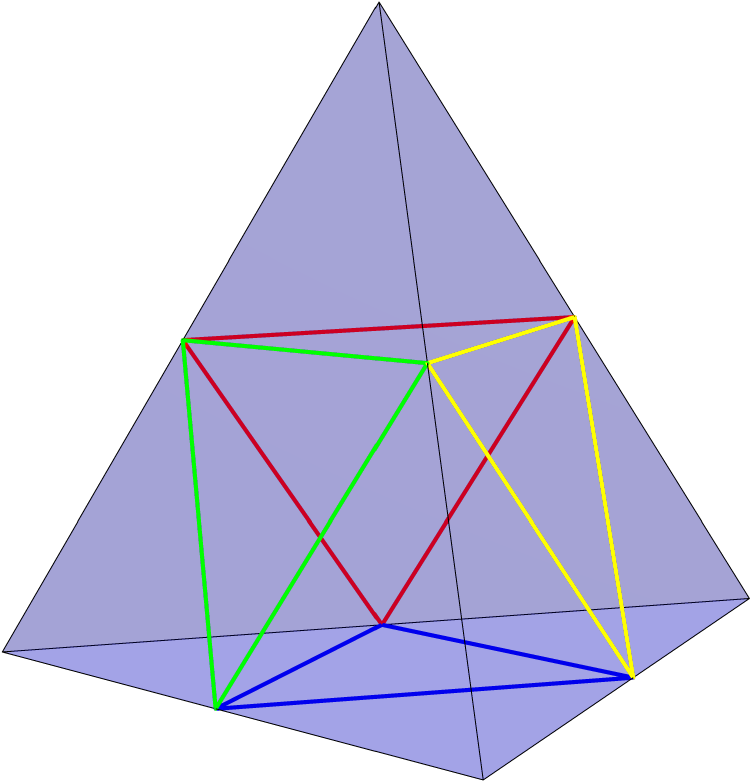}
\includegraphics[width=0.3\textwidth]{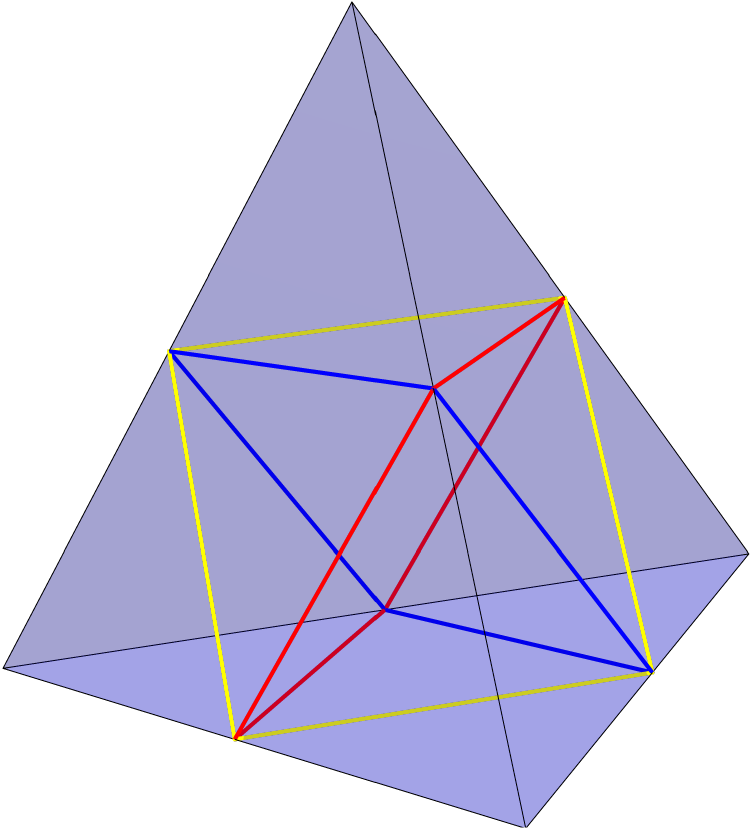}
\caption{The four triangles and the three squares that contribute to the evaluation of  $\tet$ (internal variables, by definition, are associated with their edges).} 
\label{tetrahedronT_triangles_and_squares}
\end{center}
\end{figure}

It is instructive to express the function $\tet$ in terms of the {\sl internal} triangular variables (\ie the $m,n,p$ of section sec.~\ref{admissibletriple}) rather than in terms of the {\sl external} triangular variables  (\ie the $a,b,c$ of sec.~\ref{admissibletriple}, aka highest weights of $\SU(2)$ irreps).
Calling $1,2,3,4$ the four faces of the tetrahedron $T$, 
and $m_i, n_i, p_i$ the internal triangular variables of the face $i$, we see that the semi-perimeter $\sigma(i)$ of the face $i$ is the perimeter of the triangle $(m_i,n_i,p_i)$ since  $\sigma(i)=m_i + n_i + p_i$, and that the $\tau(u)$ are the perimeters of the squares:
    \begin{equation}
        \begin{split}
\tau(1)=&\, n_1+n_2+n_3+n_4\\
\tau(2)=&\, m_1+m_2+m_3+m_4\\
\tau(3)=&\, p_1+p_2+p_3+p_4
    \end{split}
    \end{equation}
    The internal variables associated to the four triangles and the three squares contributing to the evaluation of $\tet$ can be seen in fig.~\ref{tetrahedronT_triangles_and_squares} (to be contrasted with the external variables displayed in fig.~\ref{tetrahedronT}).

\subsubsection{The $\TET$ function (also denoted $\tetrahedron_K$)}
\label{sec:TET}
In the Temperley-Lieb recoupling theory, in topological field theory, or in the theory of spin networks, one introduces the following $\TET$ function,  
{  also called tetrahedral net or tetrahedral network \cite{KauffmanLins}, chapter 8, or tetrahedron coefficient \cite{MasbaumVogel};
it can also be denoted $\tetrahedron_K$ (see sec.~\ref{sec:classicalspinnetworks})}.
{  Warning: $\TET$ is denoted Tet in  \cite{KauffmanLins}, it should not to be confused with the symbol $\tet$ used in the present paper.}
\\
Consider again a tetrahedron $T$ written for instance as $((a,b,c),(d,e,f))$.
With the same variables $\sigma(i)$,  $\tau(j)$, $m_\sigma$, $m_\tau$, as in sec.~\ref{hed_function_def}, 
one defines:

    \begin{equation}
       \TET(T)=
    \frac{J}{E} \times
   \left(
   \sum _{s=m_\sigma}^{m_\tau} 
   \frac{(-1)^s  (s+1)!}
   {\left(\prod _{i=1}^4 (s-\sigma(i))!\right)  
   \left(\prod _{j=1}^3 (\tau(j)-s)! \right)}
   \right)
    \label{TET_definition}
        \end{equation}
        
  where   \[J = \prod_{i=1,j=1}^{i=4,j=3} (\tau(j) - \sigma(j))! \qquad  \text{and} \qquad  E = {a!} {b!} {c!} {d!} {e!} {f!}\]

 {{  In \cite{KauffmanLins} or \cite{MasbaumVogel} the expression (\ref{TET_definition}) results from a calculation that we do not need to comment or repeat because in the present paper we take this expression as a definition of $\TET(T)$.}}

Obviously  $\TET(T)$, given by \ref{TET_definition},  differs from $\tet(T)$,  given by \ref{hed_definition}, by a $T$-dependent prefactor; dropping the argument $T$ we have:
    \begin{equation}
\TET = J/E \times \tet
        \end{equation}
Because of this prefactor $J/E$ the function $\TET$ is usually rational-valued but not integer-valued.
\\
For the first tetrahedron example considered in \ref{hed_function_def} we find  {\small $\TET((8, 20, 24), (15, 13, 17))$} $=\tfrac{477531}{92176448}$.

\subsubsection{From the function $\tet$ to $6j$ symbols (also denoted ${\tetrahedron}_U$)}
\label{sec:6jfromtet}
{  One potentially confusing issue is that, apart from the possible replacement of spin variables~$j$ by highest weight variables $2j$, 
 the quantities that are called ``$6j$ symbols'' in many articles of the mathematical literature (in particular in those fields related to  Temperley-Lieb recoupling theory or in the theory of invariants of $3$-manifolds),
 and also in several papers discussing topological quantum field theory or spin networks, differ from the original Wigner's $6j$ symbols used in most  physics textbooks, in \cite{KlimykSchmudgen}, or in the computation program Mathematica.
The two kinds of symbols sometimes appear in the same reference: both are for instance discussed in \cite{Carter:6J} where the first are simply called ``$6j$ symbols'' and denoted with curly braces, whereas the last are called ``normalized $6j$ symbols'' and denoted with square brackets; 
these notational conventions are not universal either since, in many textbooks (and in the present paper, see the appendix), the curly braces  denote  ``6j symbols'' that coincide with the normalized ones of  \cite{Carter:6J}.
We also adopt the in-line notation $\SixJSymbol$ for these (normalized) symbols, like in eq.~\ref{TETto6j} but the reader should remember that the arguments of the latter are spin variables, \ie half-integers, in order to agree with what is done in quantum mechanics or in Mathematica.\\
In order to ease the comparison with papers using  un-normalized $6j$ symbols, like  \cite{Carter:6J}, \cite{KauffmanLins} or \cite{MasbaumVogel}, we introduce  in eq.~\ref{TETtoRacah} the notation $\SixJSymbol_{UN}$ to denote the latter.}\\
The relation between the function $\TET$ and the $6j$ symbols of $SU(2)$ is known: {  eq.~\ref{TETtoRacah}, below, can be found for instance  in \cite{KauffmanLins} and \cite{MasbaumVogel}, and both eqs.~\ref{TETtoRacah} and \ref{TETto6j} appear in  \cite{Carter:6J}.
These last two equations give immediately a relation between $\SixJSymbol$ and $\SixJSymbol_{UN}$ but we shall not need it (this relation is also given in Lemma 2.7.10 and 2.7.12 of \cite{Carter:6J})}.
{ 
   \begin{equation}
{\SixJSymbol_{UN}}((a/2, b/2, c/2), (d/2,e/2, f/2)) =  (-1)^c (c+1) \,   \frac{\TET((a, b, c), (d,e, f))}{\ThetaGraph_K(a, b, c) \,  \ThetaGraph_K (c, d, e)} 
   \label{TETtoRacah} 
 \end{equation}
}
    \begin{equation}
    \SixJSymbol((a/2, b/2, c/2), (d/2, e/2, f/2)) =  \frac{ \TET((a, b, c), (d,e,f))}{N((a, b, c), (d,e,f))}
    \label{TETto6j}
         \end{equation}
where
   \begin{equation}N((a, b, c), (d,e,f)) =  \sqrt{ |\ThetaGraph_K(a, b, c) \ThetaGraph_K(b,d, f) \ThetaGraph_K(a,e,f) \ThetaGraph_K(c,d,e)| }   \end{equation}
This tetrahedral normalizing factor $N$ makes unitary the recoupling transformation between six spins. The presence of the square root makes irrational (in general) the values of the $6j$ symbol.
For instance,  using the previously calculated value of $\TET$ and  {\footnotesize $N((8, 20, 24), (15, 13, 17))= \frac{405 \sqrt{\frac{115}{442}}}{392}$} one obtains 
 {\footnotesize $\SixJSymbol((8/2, 20/2, 24/2), (15/2, 13/2, 17/2))=\frac{53059}{23940 \sqrt{50830}}$}.
 
\smallskip

From the relation between $\ThetaGraph_K$ and $\gon$, and from the relation between $\tet$ and $\TET$,  one finds: 
{\proposition{}  $6j$ symbols can be expressed in terms of the integer-valued functions $\gon$ and $\tet$ as follows}
   \begin{equation}
\SixJSymbol((a/2, b/2, c/2), (d/2, e/2, f/2)) = \frac{\tet((a, b, c), (d, e, f))} {\sqrt{\gon(a,b,c) \gon(b,d,f) \gon(a,e,f) \gon(c,d,e)}}
\label{tetfrom6j}
  \end{equation}    
{\footnotesize
Example: Using the values of $\gon(8, 20, 24)$, $\gon(20, 15, 17)$, $\gon(8, 13, 17)$, $\gon(24, 15, 13)$ respectively equal to
$1181079900$, $1044074631600$, $42325920$, $21903663600$, and the  value obtained for $\tet((8, 20, 24), (15, 13, 17))$ in sec.~\ref{hed_function_def},
we recover the above  $\SixJSymbol((8/2, 20/2, 24/2), (15/2, 13/2, 17/2))$}.

 The  functions $\tet$, $\TET$, and $\SixJSymbol$ can be understood as coming from different evaluation prescriptions for the same spin network $\tetrahedron$ (see the summary in sec.~\ref{sec:classicalspinnetworks}).

\subsubsection{$\tet$ and the hypergeometric function}

{  As mentioned in the appendix, an explicit expression for $6j$ symbols was obtained by Racah \cite{Racah}, as a formally infinite sum (for given arguments, only finitely many terms of this sum are nonzero) times a pre-factor involving triangular functions simply related to the function $\gon$.
From the general definition of  hypergeometric functions $_pF_q$ it was then observed (see the thesis \cite{WilsonJAthesis}, pp 29-30, and \cite{WilsonJA}) that the expression found by Racah could be written in terms of the hypergeometric function $_4F_3$. 
Formulating this last result in terms of the integer-valued functions $\tet$, see the formula \ref{hedfromhyper} below, is a simple matter of redefinition (using the relation \ref{tetfrom6j}) and of change of variables.}

\smallskip

{\bf General case.} Define 
{\footnotesize
\[ {\mathcal C} (j_ 1, j_ 2, j_ 3, j_ 4, j_ 5,  j_ 6) = \frac {(-1)^{j_ 1 + j_ 2 + j_ 4 +   j_ 5} (1 + j_ 1 + j_ 2 + j_ 4 + j_ 5)!} {(j_ 1 + j_ 2 - 
        j_ 3)! (-j_ 3 + j_ 4 + j_ 5)! (j_ 2 + j_ 4 - 
        j_ 6)! (j_ 1 + j_ 5 - j_ 6)! (-j_ 1 + j_ 3 - j_ 4 + 
        j_ 6)! (-j_ 2 + j_ 3 - j_ 5 + j_ 6)!}\]}
   and 
{\footnotesize
    \begin{equation*}
     \begin{split}  
   &  {\mathcal F}(j_1, j_2, j_3, j_4, j_5, j_6)  = \\
&  \,  _4F_3(-{j_1}-{j_2}+{j_3},{j_3}-{j_4}-{j_5},-{j_2}-{j_4}+{j_6},-{j_1}-{j_5}+{j_6};-{j_1}-{j_2}-{j_4}-{j_5}-1,-{j_1}+{
   j_3}-{j_4}+{j_6}+1,-{j_2}+{j_3}-{j_5}+{j_6}+1;1)   \end{split}   \end{equation*}
}     
where $_4F_3$ is the hypergeometric function.
{\proposition{} {  The following relation holds:}}
    \begin{equation}
\tet((a,b,c)(d,e,f))= \underset{u\to 0}{\text{lim}}  \,  {\mathcal C} (\frac{ a}{2}+ u, \frac{ b}{2}+ u, \frac{ c}{2}, \frac{ d}{2}, \frac{ e}{2}+ u, \frac{ f}{2}+ u) \times  {\mathcal F} (\frac{ a}{2}+ u, \frac{ b}{2}+ u, \frac{ c}{2}, \frac{ d}{2}, \frac{ e}{2}+ u, \frac{ f}{2}+ u)
\label{hedfromhyper}
  \end{equation}

For admissible tetrahedra the  hypergeometric contribution~${\mathcal F}$ gives a result proportional to $1/u$ when $u$ goes to $0$, whereas the coefficient ${\mathcal C}$ gives a result proportional to $u$ when $u$ goes to $0$. 
This is why we have to use a limiting procedure if we calculate $\tet$ from eq.~\ref{hedfromhyper}.
 
 \smallskip
 
 {\bf The particular values $\tet((2n,2n,2n),(2n,2n,2n))$.}
  We consider the case where all the arguments of $\tet$  (integers)  are equal, this corresponds geometrically to a regular tetrahedron. Notice that $p$ in $\tet(( p,  p,  p), ( p,  p,  p))$ cannot be odd, because the first triplet should define an admissible triangle ($3p$ should be even).
  {  From eq.~\ref{hedfromhyper} one finds:}
  {\proposition{}
Assuming that $p$ is even and setting $\tet(2n) = \tet((2 n, 2 n, 2 n), (2 n, 2 n, 2 n))$, one finds $\tet(0)=1$ and
 \begin{equation}
\tet(2n) =  \frac{(4 n + 1)!}{(n!)^4}  \, _4F_3(-n,-n,-n,-n;1,1,-4 n-1;1)
 \label{sec:regularhed}
  \end{equation}}
or, equivalently,
  \begin{equation}
\tet(2n) =  \frac{(-1)^n (1+3n)!}{(n!)^3}  \, _4F_3(-n,-n,-n,2+3n;1,1,1;1)
\tag{\ref{sec:regularhed}$'$}
 \label{sec:regularhedprime}
    \end{equation}
 This sequence starts as ${1, 96, -17010, -20160000, -5259003750, 2819345937408, \
3019973370942528, \ldots}$

{   Wilson polynomials, introduced in \cite{WilsonJA}, see also  \cite{WilsonPoly},  generalize several families of orthogonal polynomials.
Assuming $t \geq 0$, they are defined as follows in terms of the generalized hypergeometric function $_4F_3$ and the Pochhammer symbols $(u)_n=\Gamma(u+n)/\Gamma(u)$}: 
  \begin{equation}W(n, (a, b, c, d), t) = (a + b)_n  (a + c)_n  (a + d)_n  \, \, _4F_3(-n, a + b + c + d + n - 1, a - \sqrt{t}, a + \sqrt{t};  a + b, a + c, a + d; 1)  \end{equation}
  One can therefore re-write  (\ref{sec:regularhedprime}) as:
 \begin{equation}
\tet(2n) =  \frac{(-1)^n (1+3n)!}{(n!)^3} \, W(n,(-n, n + 1, n + 1, n + 1), 0)
\tag{\ref{sec:regularhed}$''$}
 \label{sec:regularhedbleprime}
  \end{equation}

\subsubsection{About speed and computational software}
 Using Mathematica  $13.0$ we compared  timings for the evaluation of
  {\small $\tet((50, 30, 76), (92, 48, 84))$ which is equal to $370574512884046997485176381045189319801237495334758378762795196256000$}
  using three different methods on a MacBook Pro (yr 2018): 
(a) From our definition of $\tet$ given by eq.~\ref{hed_definition}, (b) From eq.~\ref{tetfrom6j} and the pre-defined $\SixJSymbol$ function, (c) From eq.~\ref{hedfromhyper} and the predefined hypergeometric $_PF_Q$ function. 
The result was obtained in $0.000244\, s$ using (a), in $0.002975\, s$ using (b), and in $22.341579\, s$ using (c).
In the regular case (\ie when all the arguments are equal), one can use  eq.~\ref{sec:regularhed} rather than eq.~\ref{hedfromhyper}  and the timings become similar. 
{  Conversely, one can obtain the value of  $\SixJSymbol((j_1, j_2, j_3), (j_4, j_5, j_6))$ from eq.~\ref{tetfrom6j}, using $\tet$ defined by  eq.~\ref{hed_definition} and $\gon$ calculated from eq.~\ref{gonfromfactorial}. 
With a fresh Mathematica session, the calculation of $\SixJSymbol( \{4, 10, 12\}, \{15/2, 13/2, 17/2\})=\tfrac{53059}{23940 \sqrt{50830}}$  using the built-in $\SixJSymbol$ command  took $0.00909\, s$ (and $0.00183\, s$ if re-evaluated)  whereas it only took $0.00035\, s$ when using our functions $\gon$ and $\tet$.} 

\subsubsection{A topological identity of the $\tet$ function: the double tetrahedron identity}
\label{sec:bipyramid}

We now move to  the integer analog of the well-known Biedenharn-Elliott identity ---the latter (also called pentagon identity)  is usually written for $6j$ symbols, which, in general, are not integer valued.
{  It was originally proven \cite{Biedenharn, Elliott} by considering the coupling of four orbital momenta, and derived in \cite{BiedenharnLouck}, vol 9, pp 22-30, 
as a consequence of associativity of one of the two products that the authors define on a particular family of tensor operators. 
{  The identity written below for the integer-valued $\tet$ function is a rewriting of the Biedenharn-Elliott identity, using eq.~\ref{tetfrom6j}}. }\\
Let us consider the triangular bipyramid $B$ obtained by gluing {\sl two} tetrahedra $((b,h,k),(g,a,e))$ and  $((c,d,h),(g,e,f))$ along the common base $(e,g,h)$.
We want to associate a number $\hed(B)$ (actually an integer) to such bipyramids.

A first possibility is to define a new function $\hed_1$ as follows: 
    \begin{equation}
    \hed_1(a,b,c,d,e,f,g,h,k)=\frac{ \tet((a,b,c),(d,e,f)) \tet((a,b,c),(g,h,k))}{(-1)^{(a+b+c)/2}\gon(a,b,c)}
    \label{bipyramid1}
        \end{equation}

Intuitively, the factor $(-1)^{(a+b+c)/2} \gon(a+b+c)$ in the denominator comes from the fact that the two tetrahedra share the same triangular base.

\begin{figure}[htb]
\begin{center}
\includegraphics[width=0.3\textwidth]{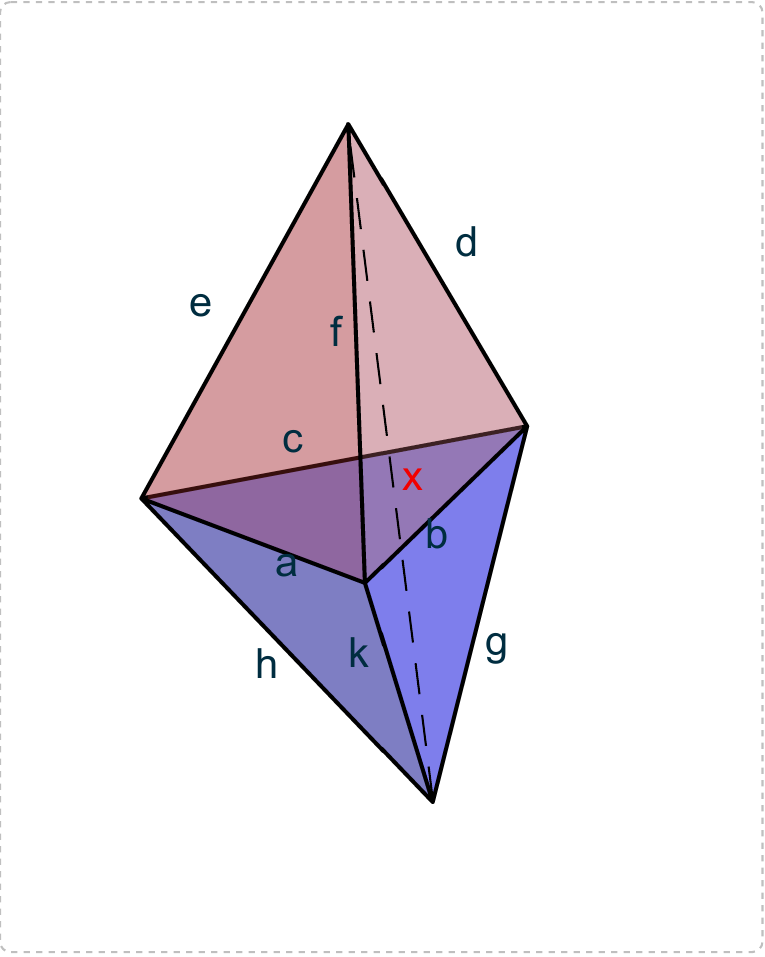}
$\qquad$
\includegraphics[width=0.3\textwidth]{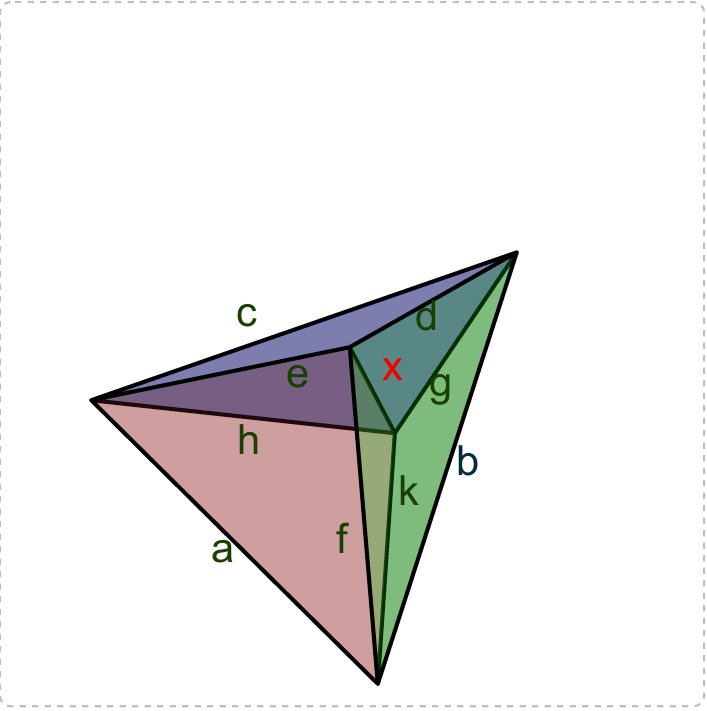}
\caption{Bipyramid. Left: as a union of two tetrahedra. Right:  as a union of three tetrahedra.} 
\label{bipyramid}
\end{center}
\end{figure}

One can however obtain the same triangular bipyramid by introducing a diagonal $x$, from ``top'' to ``bottom'', and gluing {\sl three} tetrahedra $((c,e,d),(x,g,h))$, $((b,d,f),(x,k,g))$ and $((a,e,f),(x,k,h))$.
This is illustrated in fig.~\ref{bipyramid}.
Another natural definition for a function $\hed$ would therefore be as follows:

$\hed_2(a,b,c,d,e,f,g,h,k)=$
 \begin{equation}
\sum _{x\in{\otimes[d,g] \cap \otimes[f, k] \cap \otimes[h,e]}}
\frac{
\tet((c,e,d),(x,g,h))  \tet((b,d,f),(x,k,g)) \tet((a,e,f),(x,k,h)) \, (-1)^x (x+1) }{ (-1)^{(x + d + g)/2}  \gon(x,d,g) \; (-1)^{(x+f+k)/2} \gon(x,f,k)\; (-1)^{(x+e+h)/2}\gon(x,e,h)}
     \label{bipyramid2}
        \end{equation}

The three tetrahedra appearing in $\hed_2$ share three triangles giving the factor that appear in the denominator.
The summation is performed over the common edge $(x)$,  giving a factor $(-1)^x (x+1)$ in the numerator;
this factor can be thought of as coming from the fact that the edge $(x)$ is a degenerated triangle, so that its inserted contribution is 
$(-1)^{((x + 0 + x)/2)} gon(x, 0, x)=(-1)^x (x+1)$.
   
{\proposition{One has:  $\hed_1(a,b,c,d,e,f,g,h,k)=\hed_2(a,b,c,d,e,f,g,h,k)$.}}

\smallskip
This value depends neither on the order chosen for enumerating the triangular faces of $B$ nor on  the order of the edges describing these faces.
 One can therefore define a function  $\hed(B)$ on bipyramids, either by  eq.~\ref{bipyramid1} or by eq.~\ref{bipyramid2}.

\smallskip
   
   Example.\\ Consider the bipyramid $B$ defined by {  $(a,b,c,d,e,f,g,h,k)=(28, 26, 6, 23, 19, 31, 39, 33, 17)$}\\
   From eq~\ref{bipyramid1} one finds $\hed(B)= 1395161475725373449470726604935680000$. \\
   The same value is obtained from  eq~\ref{bipyramid2}, with $x$ running in the set\\
   {  $\bigotimes[23,39] \cap \bigotimes[31,17] \cap  \bigotimes[33,19]$} $=\{16, 18, 20, 22, 24, 26, 28, 30, 32, 34, 36, 38, 40, 42, 44, 46, 48\}$.

\smallskip
 {\bf Remarks.}
 As already mentioned,  the identity (\ref{bipyramid2}) for double-tetrahedra is nothing else than the Biedenharn-Elliott identity in a new guise, since the latter is usually written in terms of $6j$ symbols which, unlike $\tet$, are not usually integer valued.
 It is tempting to try to use this ``topological identity''  (thought of as a 2-3 {  Pachner move, see \cite{Pachner_wiki}}), to obtain an integer valued function function that would coincide with $\tet$ on integer tetrahedra, with $\hed$ on integer bipyramids, and that could be generalized to arbitrary polyhedra with admissible integer sides ({  this is why the function defined by (\ref{bipyramid1}) or (\ref{bipyramid2}) was called $\hed$, as a mnemonic shortening of {\sl polyhedron}.}).
 However, trying to develop such a $3$-dimensional analog of what we did when we extended the triangle function $\gon$ to arbitrary admissible polygons does not work because, as we shall see below  (sec.~\ref{barycentricsubdivision}), when performing a barycentric subdivision, one needs to introduce an extra factor (a division by $\delta^2$ in a 1-4 Pachner move) that is not fully determined by the values attributed to the sides of the chosen polyhedron. 
 We shall say nothing in these notes about the Turaev-Viro method (that was developed in the quantum framework)  which leads to invariants of $3$-manifolds, see \cite{TuraevViro}.

In spite of the fact that the Biedenharn-Elliott identity plays an important role in the theory of spin networks, one should mention that the function $\hed$ introduced above is not associated with such a network (see sec.~\ref{sec:classicalspinnetworks}),
indeed, the 1-skeleton of the bipyramid is not a $3$-valent graph, some vertices being $4$-valent ---by way of contrast, all Wigner $3nj$ symbols are associated with (particular) $3$-valent graphs.
       
\subsubsection{Barycentric subdivision} 
\label{barycentricsubdivision}

\begin{figure}[htb]
\begin{center}
\includegraphics[width=0.6\textwidth]{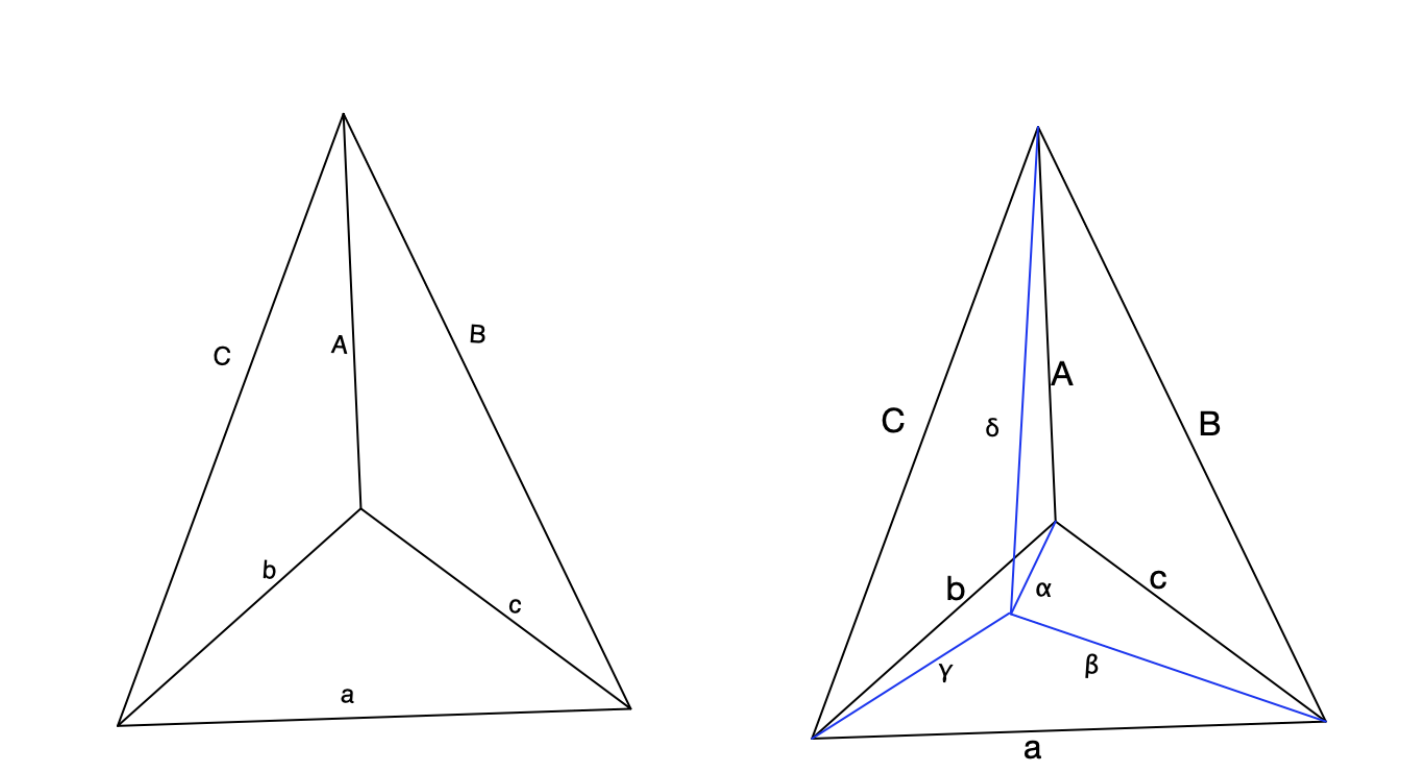}
\caption{Barycentric subdivision} 
\label{barycentric_subdivision}
\end{center}
\end{figure}

Given an admissible tetrahedron $T=\big(\begin{smallmatrix} a & b & c\\ A & B & C \end{smallmatrix}\big)$, its admissible triangles (faces) are: $(a,b,c)$, $(A,B,c)$, $(a,B,C)$, $(A,b,C)$.
Choose some point in $\R^3$ (not a vertex of  $T$); in order to obtain a geometrical interpretation of the construction we may think of that point as belonging to the inside of the tetrahedron $T$, 
it therefore defines a subdivision of the latter into a union of four tetrahedra, each one sharing three vertices of $T$ and the chosen extra vertex.
One therefore constructs the four tetrahedra:
$T_1~=~\big(\begin{smallmatrix} a & b & c\\ \alpha&\beta&\gamma \end{smallmatrix}\big)$,
$T_2=\big(\begin{smallmatrix} A & B & c\\ \beta	&\alpha&\delta \end{smallmatrix}\big)$,
$T_3=\big(\begin{smallmatrix} C & A & b\\ \alpha&\gamma&\delta \end{smallmatrix}\big)$,
$T_4=\big(\begin{smallmatrix} a & B & C\\ \delta&\gamma&\beta \end{smallmatrix}\big)$, see fig~\ref{barycentric_subdivision}.\\
Let us list the faces of all of them: \\
$T_1$ : $(a,b,c),(\alpha,\beta,c),(a,\beta,\gamma),(\alpha,b,\gamma)$,  
$T_2$ : $(A,B,c),(\beta,\alpha,c),(A,\alpha,\delta),(\beta,B,\delta)$,\\
$T_3$ : $(C,A,b),(\alpha,\gamma,b),(C,\gamma,\delta),(\alpha,A,\delta)$,
$T_4$ : $(a,B,C),(\delta,\gamma,c),(a,\gamma,\beta),(\delta,B,\beta)$.

Among the admissible triangles of the tetrahedra $T_1, T_2, T_3, T_4$, some are ``already known'' because they are also faces of $T$;
notice also that some triangles appear twice in the above lists. For all the $T_i$'s to be admissible one needs to impose admissibility on the following six new faces:
$\{(\alpha,\beta,c)$, $(a,\beta,\gamma)$, $(\alpha,b,\gamma)$, $(A,\alpha,\delta)$, $(\beta,B,\delta)$, $(\gamma,\delta,C)\}$.
Given an admissible tetrahedron $T$ there is an infinite number of possible barycentric subdivisions defined by a quadruple $(\alpha, \beta, \gamma, \delta)$, however, if one internal edge (for instance $\delta$) is fixed to some integer value, 
 there is only a finite set of triplets $(\alpha, \beta, \gamma)$ for which the above six triangular conditions will be obeyed.
 The cardinality of this set of triplets depends of course on the given tetrahedra $T$; it increases with the value of the chosen $\delta$ and stabilizes when $\delta$ reaches some value $\delta_{\text{max}}$.\\
 Let us illustrate this comment with the following example:

Take $T=((2, 1, 3), (1, 2, 2))$.  Then, for $\delta=0,1,2,3,4,5,\ldots$ one obtains the following possibilities for the triplets $(\alpha, \beta, \gamma)$:
{\footnotesize
\[\left(
\begin{array}{ccc}
 1 & 2 & 2 \\
\end{array}
\right),\left(
\begin{array}{ccc}
 0 & 3 & 1 \\
 2 & 1 & 1 \\
 2 & 1 & 3 \\
 2 & 3 & 1 \\
 2 & 3 & 3 \\
\end{array}
\right),\left(
\begin{array}{ccc}
 1 & 2 & 0 \\
 1 & 2 & 2 \\
 1 & 4 & 2 \\
 3 & 0 & 2 \\
 3 & 2 & 2 \\
 3 & 2 & 4 \\
 3 & 4 & 2 \\
 3 & 4 & 4 \\
\end{array}
\right),\left(
\begin{array}{ccc}
 2 & 1 & 1 \\
 2 & 1 & 3 \\
 2 & 3 & 1 \\
 2 & 3 & 3 \\
 2 & 5 & 3 \\
 4 & 1 & 3 \\
 4 & 3 & 3 \\
 4 & 3 & 5 \\
 4 & 5 & 3 \\
 4 & 5 & 5 \\
\end{array}
\right),\left(
\begin{array}{ccc}
 3 & 2 & 2 \\
 3 & 2 & 4 \\
 3 & 4 & 2 \\
 3 & 4 & 4 \\
 3 & 6 & 4 \\
 5 & 2 & 4 \\
 5 & 4 & 4 \\
 5 & 4 & 6 \\
 5 & 6 & 4 \\
 5 & 6 & 6 \\
\end{array}
\right)
,
\left(
\begin{array}{ccc}
 4 & 3 & 3 \\
 4 & 3 & 5 \\
 4 & 5 & 3 \\
 4 & 5 & 5 \\
 4 & 7 & 5 \\
 6 & 3 & 5 \\
 6 & 5 & 5 \\
 6 & 5 & 7 \\
 6 & 7 & 5 \\
 6 & 7 & 7 \\
\end{array}
\right)
, \ldots
\]}
The cardinality of the set of solutions is $1,5,8,10,10,10,\ldots$ and stabilizes at $10$  when $\delta \geq \delta_{\text{max}}=3$.

Given $T$, choosing $\delta$ amounts to choose a finite family of subdivisions (triangulations) of $T$.  
Since each subdivision involves four tetrahedra sharing six faces, and since these faces share four new edges $\alpha, \beta, \gamma, \delta$, 
one is led to consider the expression: 
{\definition
 {\small  
 \begin{equation}
 \begin{split}
{} &P(\delta) = \sum_{\alpha, \beta, \gamma}  \, P_1 P_2 P_3  \\
  \text{where}&{} \\
 P_1 &=\frac{
  \tet((a,b,c),(\alpha ,\beta ,\gamma ))\tet((a,B,C),(\delta ,\gamma ,\beta ))\tet((C,A,b),(\alpha ,\gamma
   ,\delta ))\tet((A,B,c),(\beta ,\alpha ,\delta ))}{\text{gon}(a,\beta ,\gamma )\gon(A,\alpha ,\delta )\gon(\alpha
   ,b,\gamma )\gon(\beta ,B,\delta )\gon(\alpha ,\beta ,c)\gon(\gamma ,\delta ,C)}\\
  P_2 &=(-1)^\alpha (\alpha + 1) (-1)^\beta (\beta + 1) (-1)^\gamma   (\gamma + 1) (-1)^\delta (\delta + 1) \\
   P_3&=(-1)^{\frac{1}{2} (a+\beta +\gamma )} (-1)^{\frac{1}{2} (\alpha +A+\delta )} (-1)^{\frac{1}{2} (\alpha +b+\gamma )}
   (-1)^{\frac{1}{2} (\beta +B+\delta )} (-1)^{\frac{1}{2} (\alpha +\beta +c)} (-1)^{\frac{1}{2} (\gamma +C-\delta )}
   \end{split}
   \label{barycentric}
   \end{equation}
  }
The sum $\sum_{\alpha, \beta, \gamma}$ runs over all the triplets $(\alpha, \beta, \gamma)$ that make admissible the six triangles defined by the choice of the edge $\delta$ ---see the previous discussion.}
   
{\proposition{}   $\frac{1}{(\delta+1)^2} \, P(\delta)$ is independent of $\delta$, and,  
 \begin{equation} 
 \forall{\delta \in \N},  \quad \frac{1}{(\delta+1)^2} \, P(\delta) =  \tet((a,b,c)(A,B,C))
 \label{eq:barycentric}
    \end{equation}}

Let us illustrate this property with the example previously considered. \\
One finds $\tet((2, 1, 3), (1, 2, 2))=-24$ from the definition of $\tet$.
Then, for the triplets $(\alpha, \beta, \gamma)$ previously displayed, which are associated with successive choices of $\delta=0,1,2,3,4,5,\ldots$, one obtains the following contributions to the sum  $\sum_{\alpha, \beta, \gamma}$ of eq.~\ref{barycentric}:
 \[
\begin{array}{c}
 \left(
\begin{array}{c}
 -24 \\
\end{array}
\right),
 \left(
\begin{array}{c}
 -24 \\
 -\frac{64}{3} \\
 -\frac{32}{3} \\
 \frac{40}{3} \\
 -\frac{160}{3} \\
\end{array}
\right), 
 \left(
\begin{array}{c}
 -24 \\
 12 \\
 -60 \\
 -24 \\
 -30 \\
 -30 \\
 30 \\
 -90 \\
\end{array}
\right),
 \left(
\begin{array}{c}
 -\frac{32}{3} \\
 \frac{20}{3} \\
 -\frac{160}{3} \\
 \frac{64}{3} \\
 -108 \\
 -60 \\
 -\frac{192}{5} \\
 -\frac{288}{5} \\
 \frac{252}{5} \\
 -\frac{672}{5} \\
\end{array}
\right),
 \left(
\begin{array}{c}
 -30 \\
 18 \\
 -90 \\
 30 \\
 -168 \\
 -108 \\
 -\frac{140}{3} \\
 -\frac{280}{3} \\
 \frac{224}{3} \\
 -\frac{560}{3} \\
\end{array}
\right),
 \left(
\begin{array}{c}
 -\frac{288}{5} \\
 \frac{168}{5} \\
 -\frac{672}{5} \\
 \frac{192}{5} \\
 -240 \\
 -168 \\
 -\frac{384}{7} \\
 -\frac{960}{7} \\
 \frac{720}{7} \\
 -\frac{1728}{7} \\
\end{array}
\right), \ldots 
\end{array}
\]

Summing over the columns gives the following values for  $P(\delta) = {-24, -96, -216, -384, -600, -864}$.
Dividing these sums by $1/(\delta+1)^2$ gives back the constant value $-24$.

\subsubsection{Playing with a cube}
\label{tetTOhedAndCube}

In this subsection we consider a cube (actually a parallelepiped)  decorated with a family of compatible non-negative integers,  to which we attach a quantity, that we call $cube$, using some kind of statistical sum involving the integer-valued functions $\gon$ and $\tet$ previously discussed.
This quantity does not seem to be related with standard constructions (in particular it is not an integer variant of the cube function that one knows how to define in the theory of spin networks); 
as a matter of fact this subsection is not very much related to the other topics discussed in these set of notes,  and we could (maybe should) have removed it from the manuscript. 
It remains, however, that we find $cube$ interesting, and we hope that some reader will find a way to incorporate this quantity naturally in a general framework, or  find some use for it.

\begin{figure}[htb]
\begin{center}
\includegraphics[width=0.4\textwidth]{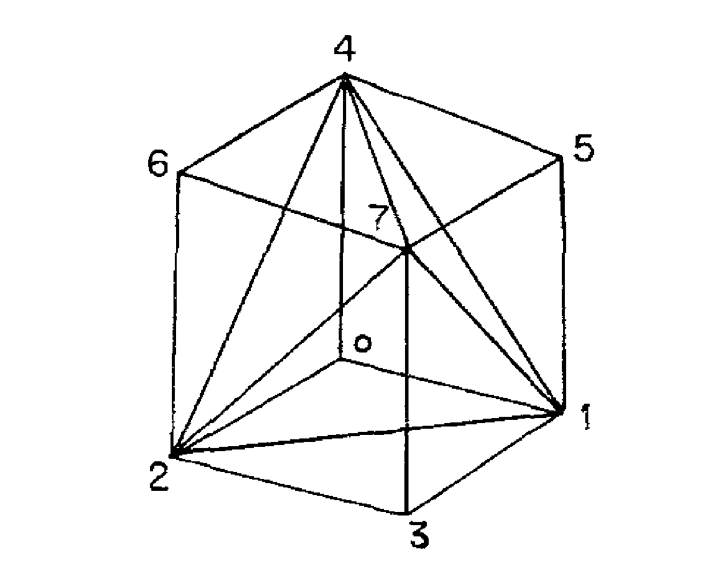}
\caption{Minimal triangulation of the cube} 
\label{minimal_triangulation_cube}
\end{center}
\end{figure}

Consider a cube in $\R^3$.
Its standard triangulation yields six $3$-simplices (six congruent birectangular tetrahedra);
however one can find a triangulation containing only five $3$-simplices: slice off four corners (not edge-related), this makes four congruent three-right-angled tetrahedra, and what is left inside is a regular tetrahedron, see fig.~\ref{minimal_triangulation_cube}.
As it is well known, $5$ is the minimal number of simplices in a triangulation of a $3$-cube.
Looking at fig.~\ref{minimal_triangulation_cube} we see that the following simplices (tetrahedra) are formed: 
$[0,1,2,4]$,
$[1,2,3,7]$,
$[2,4,6,7]$,
$[1,4,5,7]$, and 
$[1,2,4,7]$.
With the previous labels to denote vertices we use pairs of labels to denote the edges (and a natural ordering to avoid double counting).
In terms of edges we have the following sub-tetrahedra: $((01,12,20)(24,04,14))$, $((24,46,62),(67,27,47))$, $((14,45,51),(57,17,47))$, $((12,23,31),(37,17,27))$, $((12,24,41),(47,17,27))$.
Only the last one is regular, the others are rectangular. There are four shared faces that build the regular tetrahedron, with six common edges, 12, 14, 17, 24, 27, 47, and four common points: 1, 2, 4, 7.
The 12 given cube edges are
01, 13, 32, 20
45, 57, 76, 64
04, 15, 26, 37.
The new edges appearing in the triangulation are  6 diagonals (the edges of the regular tetrahedron), namely:
14, 17, 27, 24, 12, 47.

An interesting cube function could be defined as follows.
We suppose given 12 non-negative integers for the edges  of the cube, actually a parallelepiped. 
We first determine all the sets of  four rectangular tetrahedra defining a possible dissection by imposing admissibility conditions on the tetrahedra. The number of such sets is finite.
For each such possibility we take the product of the functions $\tet$ for the five tetrahedra of the dissection, 
we then divide by the $\gon$ functions associated with the four common faces (triangles) of the fifth tetrahedron,  and multiply the obtained result by the contribution $(x_{ij}+1)$ of the
six new edges.
Finally we sum over these possibilities, each of them corresponding to the six new edges that appear in the dissection (the diagonals of the faces of the cube).
\\
Examples.
Assuming that all 12 edges are equal to $n$, we call $cube(n)$ the result. One finds $cube(0)=1$, 
$cube(1)=-63488$ (a sum over 15 possibilities), 
$cube(2)=+5580307647$ (a sum over 127 possibilities), 
$cube(3)=-297180797599744$ (a sum over 648 possibilities).\\
One may of course consider parallelepipeds with unequal edges. 
For instance one can choose the edges $x_{01}= 2$, $x_{13}= 1$, $x_{23}= 2$, $x_{02}= 1$, $x_{45}= 2$, $x_{57}= 1$, $x_{67}= 2$, 
 $x_{46}= 1$, $x_{04}= 1$, $x_{15}= 1$, $x_{26}= 1$, $x_{37}= 1$, and obtain $1994112$ as the value for the corresponding ``$cube$'' function . 

The above ``$cube$ function'' is not a priori related to the cube function that one can define as a spin network (evaluation of the latter makes sense, see sec.~\ref{sec:classicalspinnetworks},  since the 1-skeleton of the $3$-dimensional cube is indeed a trivalent graph).
For a cube with all edges equal to $n$, the latter, and its asymptotics, were considered by Don Zagier (appendix of \cite{GaroufalidisEtAl}).  Its values for the first values of $n$ are:  $1$, $6144$, $505197000$,  $77414400000000$, $\ldots$

\subsubsection{Other properties of the function $\tet$}
\label{otherpropertiesoftet}
 
  \smallskip    
{\bf Regge symmetry of the function $\tet$.}
Let us consider the four semi-perimeters of the non-planar quadrilaterals defined by a tetrahedron (see fig.~\ref{tetrahedronT}): $p_{12} = (a + b + d + e)/2$, $p_{23} = (b + c + e + f)/2$, $p_{31} = (a + c + d + f)/2$.
{\proposition{}
For each choice of such a quadrilateral, the $\tet$ function is invariant when its arguments, that are also edges entering the chosen quadrilateral, are simultaneously shifted as follows by the corresponding semi-perimeter:
\begin{equation}
\begin{split}
\tet((a, b, c), (d, e, f)) =&  \tet((p_{12} - a, p_{12} - b, c), (p_{12} - d, p_{12} - e, f))\\
\tet((a, b, c), (d, e, f)) =& \tet((a, p_{23} - b, p_{23} - c), (d, p_{23} - e, p_{23} - f))\\
\tet((a, b, c), (d, e, f)) =& \tet((p_{31} - a, b, p_{31} - c), (p_{31} - d, e, p_{31} - f))
\end{split}
\label{ReggeSymmetry}
\end{equation}}
Proof: Using the relation between the function $\tet$ and the $6j$ symbols  (see eq.~{\ref{tetfrom6j}), one obtains the above property from the fact that the latter are known to be invariant under the same transformation (Regge symmetry).
For a modern treatment of these symmetries, see \cite{Boalch}.

 \smallskip      
{\bf Asymptotics of the function $\tet$.}
{  Expanding on the work of Wigner \cite{Wigner}, an asymptotic formula relating the value of the $6j$-symbol, when the dimensions of the representations are large, to the volume of an Euclidean tetrahedron whose edge lengths are these dimensions, was conjectured in \cite{PonzanoRegge} (Ponzano-Regge formula). 
The proof, together with an interpretation in terms of geometric quantization, was obtained in \cite{Roberts:tetrahedron}, see also \cite{Roberts:quantumtetrahedron}.  
This formula assumes that all the arguments are large but it is also interesting to study what happens when one or more arguments stay constant whereas the others increase. One may picture the various asymptotic regimes in terms of vertices or in terms of edges of an associated tetrahedron, and the various asymptotic regimes may be called $[1,1,1,1]$ (all the edges are large), $[1,1,2]$ (one edge stay small), $[2,2]$ (two edges stay small) and $[1,3]$ (three edges stay small).
The $[1,3]$ case was explicitly considered in one section of \cite{PonzanoRegge}, where the  formula below, eq.~\ref{6JAsymptotics31}, was proved (this expression had already been mentioned in \cite{Adleretal}).
Sketch of the proof: express the $6j$ symbols as a sum of products of four $3j$ symbols (as before), write the $3j$'s in terms of factorials, and use the Stirling formula to study the chosen asymptotic regime.}
\\ For given compatible arguments $a,b,c,d,e,f$, the asymptotic equivalent, 
when $r$ goes to infinity, of a $r$-shifted and rescaled $6j$ symbol can be written in terms of a Wigner $3j$ symbol as follows: 

  {\footnotesize
 \begin{equation}
 \sqrt{(2 r)} \, \SixJSymbol((a, b, c), (d + r, e + r, f + r))  \sim_{r \rightarrow \infty}(-1)^{(a + b + c + 2 (d + e + f))} \text{ThreeJSymbol}((a, e - f), (b,  f - d), (c, d - e)).
 \label{6JAsymptotics31}
  \end{equation}}
  
 {  The interested reader can} obtain an expression giving an asymptotic equivalent for the $\tet$ function in terms of  one $3j$ symbol  (or of one Clebsch-Gordan coefficient) 
 and several appropriate $\gon$ factors from eq.~\ref{6JAsymptotics31}  by expressing the $6j$ symbol in terms of the $\tet$ function using eq.~\ref{tetfrom6j}.

\smallskip
\paragraph {{  Recoupling coefficients}, Racah coefficients,  cells, and orthogonality.}
Let $T$ be an admissible tetrahedra labelled with integers $a,b,c,d,e,f$ (twice the spin variables),  as in fig.~\ref{tetrahedronT}.
Choose a pair of opposite edges, the two other pairs of opposite  edges building a closed (and skew) quadrilateral. Obviously there are three possible choices.
For instance we may choose {  the pair carrying the labels $(b,e)$}, the corresponding quadrilateral being $(a,c,d,f)$; also denoted  $\big(\begin{smallmatrix} a & c \\ d & f\end{smallmatrix}\big)$: 
{  This quadrilateral is obtained by removing one column in the symbol specifying the tetrahedron $T$, the position of this column in the symbol is actually not relevant: it may be the second column if we denote $T$ by $\big(\begin{smallmatrix} a & b & c\\ d & e & f \end{smallmatrix}\big)$
but it will be for instance the third column if we denote the same tetrahedron by $\big(\begin{smallmatrix} a & c & b\\ d & f & e \end{smallmatrix}\big)$.}
For the choice $(b,e)$, a quantity of interest is
    \begin{equation}
    U_{be}=(-1)^{(a+c+d+f)/2}  \sqrt{(b+1)(e+1)}  \, \SixJSymbol(T).
    \label{recouplingfrom6j}
        \end{equation}
It is proportional to the $6j$ symbol but the removed column gives an extra square root, and the left-over quadrilateral gives a sign. 
{    The choice of another pair of opposite edges in $T$ leads to similar coefficients $U_{be}$, $U_{ad}$, $U_{cf}$.
From now on, for definiteness,  we choose the edge $(b,e)$.
This number is called a {\sl recoupling coefficient} \cite{Racahcoefficient}, or  a {\sl unitary Racah coefficient}  (see \cite{BiedenharnLouck}, vol 1,  p\,111, or \cite{Hecht}, p\,314).
The Racah coefficient itself is $W_{be} = (-1)^{(a+c+d+f)/2}  \SixJSymbol(T)$, it  differs from $U_{be}$ by a square root and from the associated $6j$ symbol by a possible sign.
Another notation used in the literature for the same Racah coefficient, using spin variables and displaying the six arguments, is $W(a/2,c/2,f/2,d/2;\,b/2,e/2)$ but the order of arguments is not unique because of the tetrahedral symmetries of $T$.
Racah coefficients were defined in \cite{Racah}, see also the discussion in \cite{BiedenharnLouck}, vol 1, p\,101, or in \cite{Racahcoefficient}. 
We only mention them here in order to avoid confusion with similar quantities, indeed it is not uncommon to see the name of Racah associated with distinct families of objects (the $6j$'s, the un-normalized version of the $6j$'s, the recoupling coefficients, or the Racah coefficients themselves).

In terms of the two integer-valued  functions $\gon$ and $\tet$, the recoupling coefficients read as follows (using eqs.~\ref{recouplingfrom6j} and  \ref{tetfrom6j}):
    \begin{equation} 
    U_{be} = \frac{(-1)^{(a+c+d+f)/2} \sqrt{(b + 1) (e + 1)} \, \tet[(a, b, c), (d, e, f)]}  {\sqrt{\gon(a, b, c) \gon(a, e, f) \gon(b, d, f) \gon(c, d, e)}} \label{recouplingfromtet}
  \end{equation}

{\proposition  For given non-negative integers ${a,c,d,f}$, consider the matrix $U$ with matrix elements $U_{be}$ as in  \ref{recouplingfromtet}, where $(b,e)$ belongs to the set of pairs of non-negative integers for which the tetrahedron $T=((a, b, c), (d, e, f))$ is admissible.  
Then we have the following proposition: The matrix $U$ is orthogonal i.e. $U.U^T = U^T.U = 1$. } \\
This is nothing but a rewriting of the orthogonality property for the recoupling coefficients, which itself can be traced back to a similar property (discussed in many places) for the associated $6j$ symbols.}

Remark: The quantities $U$ may also be called ``cells'' because they describe the pairing defining a bigebra structure (actually a  weak Hopf algebra) on the vector space ${\mathcal B}$ of double triangles; this is a particular case of a much more general construction (``Ocneanu cells'').
In a nutshell, one considers formally the graded vector space generated by admissible triplets $(a,b,c)$ pictured as triangles (the grading is defined by $b$, which is displayed horizontally), and its corresponding graded endomorphism algebra ${\mathcal B}$.
A particular basis of the latter is conveniently described in terms of ``horizontal double triangles'': an horizontal double triangle of type $b$ is a pair of two triangles sharing the same edge $b$, displayed horizontally.
Dually we have also ``vertical double triangles'' (spanning the dual of ${\mathcal B}$) where the common edges are displayed vertically. The numbers $U$ define a non trivial pairing: they are used to define a coalgebra structure on ${\mathcal B}$. 
The obtained bigebra is infinite dimensional since the grading $b$ ranges over $\N$.
The theory has been developed in the quantum case ({  when $q$ is a root of unity,}  ${\mathcal B}$ is finite dimensional); a brief account is given in \cite{Coquereaux:cells}, which also contains a study of several explicit examples.
{  In the classical case, the existence and properties of the two algebra structures was discussed in  \cite{BiedenharnLouck}, vol 2, using another language and notations.}
In the present notes we are mostly interested in the arithmetic properties of the integer-valued functions $\gon$ and $\tet$, so this is not the right place to discuss the above in more details.

\smallskip
{\bf Expressing the function $\tet$ in terms of Clebsch-Gordan coefficients.}
Since the integer valued function $\tet$  differs by normalization factors only from the $6j$ symbol, see eq.~\ref{tetfrom6j}, {  and since the latter can be written as a multiple sum over a product of Clebsch-Gordan coefficients (or of $3j$ symbols), see the Appendix (sec.~\ref{appendix}),}
it is clear that one can also express  $\tet$ as a sum involving products of Clebsch-Gordan coefficients.
This is left as an exercise.
   
\section{Quantum version}
{  The notions and properties discussed in the previous section have a $q$-analog counterpart. In the present section we shall only give the corresponding definitions, state the analogous properties, and, only in a few cases, provide sketch of proofs.}

\subsection{$q$-numbers and $q$-factorials}
If $n$ is a real number, in particular if it is a positive integer, and $q$ is complex, there are two variants of $q$-numbers that we recall below.
 \begin{equation} [n]_q = \frac{q^n - q^{-n}}{q-q^{-1}} \quad \text{and}  \quad [[n]]_q =  \frac{1 - q^{n}}{1-q} \end{equation}
Here we only use the first convention, and sometimes ommit the surrounding square bracket, setting $n_q = ~ [n]_q$.
There are also two variants of $q$-factorials, that of course differ by prefactors: $[s] !_q = \prod_{n=1}^{n=s} [n]_q$ and $[[s]] !_q = \prod_{n=1}^{n=s} [[n]]_q$.
They are related as follows: $[[s]]!_{q^2} = q^{n(n-1)/2} \; [s]!_{q}$. Warning: the pre-defined function $\text{QFactorial}[s, q]$ of Mathematica, which is displayed in this program (using the so called``traditional form'') as $[s]_q!$,  coincides with our $[[s]] !_q$.
We shall only use the first notion of $q$-factorial (\ie our $[s]_q!$) in the following.

\subsection{The quantum $\gon_q$ function}
\subsubsection{$q$-admissible triplets}
\label{qadmissibility}

$q$-admissibility.   If $q$ is a root of unity with $q = \exp(i\pi/\kappa)$, with $\kappa$ an integer larger or equal to $2$, so that $q^\kappa=-1$, we do not only assume that the triplet $(a,b,c)$ is admissible in  the classical sense, \ie triangular and $a+b+c$ even, but we also assume that $a + b + c \leq 2\kappa-4$.
It is standard to call ``level'' the integer $\ell=\kappa-2$.

\subsubsection{The elementary quantum $\gon_q$ function  (triangular function)}
The quantum $q$-triangle function (\ie the $\gon_q$ function with three arguments) is defined as follows:
{\definition
 \begin{equation}{\gon_q}(a,b,c)=\frac{\left[\frac{1}{2} (a+b+c)+1\right]!_q}{\left[\frac{1}{2} (a+b-c)\right]!_q \left[\frac{1}{2} (a-b+c)\right]!_q \left[\frac{1}{2} (-a+b+c)\right]!_q}\end{equation}
 When $q$ is a root of unity we assume that $(a,b,c)$ is $q$-admissible, otherwise we set ${\gon_q}(a,b,c)=0$.}

\smallskip
Example:\\ $\gon_q(3,7,8)=\frac{7_q \, 8_q\,  9_q \, 10_q}{1_q^2 \, 2_q}$,  whose classical limit (\ie $q \rightarrow 1$) is $\gon(3,7,8)=2520$, explicitly reads:
{\tiny  \[\frac{\left(q^{12}+q^8+q^4+1\right) \left(q^{12}+q^{10}+q^8+q^6+q^4+q^2+1\right)
   \left(q^{16}+q^{14}+q^{12}+q^{10}+q^8+q^6+q^4+q^2+1\right)
   \left(q^{18}+q^{16}+q^{14}+q^{12}+q^{10}+q^8+q^6+q^4+q^2+1\right)}{q^{29}}\]}

\subsubsection{Duality property: a topological identity of the $\gon_q$ function.}
{\proposition Like in the classical case we have the duality property:
\begin{equation} \label{qdualityidentity}
\sum_{s\in S} \gon_q(a,b,s) \frac{1}{[s+1]_q} \gon_q(c,d,s) = \sum_{t \in T} \gon_q(a,d,t) \frac{1}{[t+1]_q} \gon_q(b,c,t)  \end{equation}
 where $S$ (resp. $T$) denote the set of integers $s$ (resp. $t$) making $q$-admissible both triplets $(a,b,s)$ and $(c,d,s)$  (resp. both $(a,d,t)$ and $(b,c,t)$).\\
The above sums are also equal to $\sum_{u \in U} \gon_q(a,c,u) \frac{1}{[u+1]_q} \gon_q(b,d,u)$ where $u$ runs over the set $U$ of integers making $q$-admissible both  $(a,c,u)$ and $(b,d,u)$.
Any one of these three sums defines the same $q$-quadrilateral function $\gon_q(a,b,c,d)$.}
 
{   As in the classical case, the identity $\ref{qdualityidentity}$  comes from an isomorphism between the vector spaces of intertwiners $Hom_{SU(2)_q} (V_a \otimes V_b,  V_c \otimes V_d)$ or
$Hom_{SU(2)_q} (V_a \otimes V_d,  V_b \otimes V_c)$ and $Hom_{SU(2)_q} (V_a \otimes V_b \otimes V_c \otimes V_d, \C)$.  
The word ``duality'' comes from the Frobenius duality (or reciprocity) briefly recalled in our discussion in \ref{Hom-spaces}.  We are not aware of any published purely $q$-deformed combinatorial proof of this identity.}

  \subsubsection{Extension of the function $\gon_q$ to arbitrary $q$-admissible multisets}
  Like in the classical case, the above topological property allows one to extend the definition of $\gon_q$ to arbitrary $q$-admissible multisets, by choosing some triangulation of any polygon whose edges are elements of this multiset.
One takes the  product of the triangular $\gon_q$ functions over all the 2-simplices of the triangulation, divided by a product of factors $[x+1]_q$ where the $x$'s  are the lengths of the diagonals (1-simplices) entering the chosen triangulation.
  Property (\ref{qdualityidentity}) ensures that the result is independent of the choices. 
  In the root of unity case, $q$-admissibility can be re-phrased in terms of representation theory of the quantum group $SU_q(2)$ by imposing that the product of the simple objects defined by the arguments (in the associated appropriate monoidal category) contain the trivial representation.
  
The above leads to the recursive definition:
{\definition
\label{qgondefinition}
\begin{equation}
\gon_q(x_1,x_2,\ldots,x_n,u,v) = \sum_x \gon_q(x_1,x_2,\ldots,x_n,x) \, \frac{1}{[x+1]_q} \, \gon_q(x,u,v)\end{equation}
where the sums runs over elements $x$ of the intersection {  of the underlying sets associated} with the multisets {$\otimes[x_1,x_2,\ldots,x_n]$ and $\otimes[u, v]$}.}\\
As before the notation $\bigotimes[a_i]$ refers to the multiset of highest weights that appear in the direct sum decomposition  (irreps) of the chosen tensor products.
{  In the root of unity case, the determination of $\bigotimes[a_i]$ is quite subtle because some of the multiplicities can be smaller than in the classical case if the chosen level is not chosen big enough, and some of the highest weight classically present in the decomposition can even disappear.
We shall not discuss that issue further in the present paper since it is discussed at length in articles devoted to the study of quantum groups or of conformal field theory.
As in the classical case, the symmetry property of $\gon_q$ comes from the previous double triangle identity \ref{qdualityidentity} and from the symmetry property of the elementary $gon_q$ function.}

  \smallskip
 Example:  We look at the pentagonal example defined by the set  $\{11, 3, 4, 1, 5\}$ that we already considered in the classical case.
 If $q$ is a root of unity we assume that we choose a level big enough in order not to worry with the possible disappearance of some representations.
  Using any triangulation one finds $\gon_q[11, 3, 4, 1, 5]=\frac{6_q 7_q 8_q 9_q 10_q 11_q 12_q \left(1_q 5_q 6_q+10_q \left(4_q 6_q+3_q 13_q\right)\right)}{1_q^4 2_q^2 3_q^2 4_q}$.\\
 Explicitly, this reads:
 {\tiny
\begin{equation*}
\begin{split}
 q^{-70} & \left(q^4-q^2+1\right)^2 \left(q^8+1\right) \left(q^8-q^4+1\right)
   \left(q^8-q^6+q^4-q^2+1\right) \left(q^8+q^6+q^4+q^2+1\right)^2 \left(q^{10}+2 q^8+3
   q^6+3 q^4+2 q^2+1\right)^2 \left(q^{12}+q^6+1\right) \\
 &  \left(q^{12}+q^{10}+q^8+q^6+q^4+q^2+1\right)
   \left(q^{20}+q^{18}+q^{16}+q^{14}+q^{12}+q^{10}+q^8+q^6+q^4+q^2+1\right)\\
 &  \left(q^{28}+q^{22}+q^{20}+q^{18}+q^{16}+q^{14}+q^{12}+q^{10}+q^8+q^6+1\right)
   \end{split}
   \end{equation*}
   }
   The reader can check that the classical limit ($q\rightarrow 1$) of this expression is equal to $18295200$, as it should (see sec.~\ref{generalpolygon}).
     After multiplication by $q$ raised to some appropriate power ($70$ in the above example) we obtain a polynomial in $q$. It is actually monic symmetric unimodal in the variable $q^2$.
   
{  \subsubsection{The $\ThetaGraph_{K,q}$ function}
\label{KauffmanThetaq}
The $q$-analog of the classical theta net whose definition was recalled in section \ref{KauffmanTheta} is discussed in many places, see for instance  \cite{KauffmanLins} , \cite{Carter:6J}.
Its expression is given by the quantum analog of eq.~\ref{classicalthetadef}, the factorials being replaced by $q$-factorials: 
\begin{equation}
\ThetaGraph_{K,q} (a,b,c)  = (-1)^{m+n+p}    \frac{[m+n+p+1]!_q \,  [m]!_q [n]!_q  [p]!_q }{[m+n]!_q [m+p]!_q [n+p]!_q} \label{quantumthetadef} 
\end{equation} 
Its relation with the elementary $\gon_q$ function is given by the $q$-analog of eq.~\ref{gontotheta}.
 }

\subsubsection{The function $\gon_q$ and quantum Hilbert matrices}
The $q$-Hilbert matrix $H_q$ of order $n$ is defined as the $n\times n$ matrix $H_q(n)$ with matrix elements  $H_q(n)(i,j)= 1/[i +j-1]_q$.

\smallskip
For instance
$H_q(3)=\left(
\begin{array}{ccc}
 \frac{1}{[1]_q} & \frac{1}{[2]_q} & \frac{1}{[3]_q} \\
 \frac{1}{[2]_q} & \frac{1}{[3]_q} & \frac{1}{[4]_q} \\
 \frac{1}{[3]_q} & \frac{1}{[4]_q} & \frac{1}{[5]_q} \\
\end{array}
\right) =
\left(
\begin{array}{ccc}
 1 & \frac{q}{q^2+1} & \frac{q^2}{q^4+q^2+1} \\
 \frac{q}{q^2+1} & \frac{q^2}{q^4+q^2+1} & \frac{q^3}{q^6+q^4+q^2+1} \\
 \frac{q^2}{q^4+q^2+1} & \frac{q^3}{q^6+q^4+q^2+1} & \frac{q^4}{q^8+q^6+q^4+q^2+1} \\
\end{array}
\right).$
  
   Several formulae have been obtained in the literature  for the entries of the inverse Hilbert matrix, classical (see for instance \cite{HilbertMatrix_wiki}),  or quantum (see \cite{AndersenBerg}).
   {  The corresponding trace, calculated for instance using eq.~33 of \cite{AndersenBerg}, gives the same result as $\gon_q(n,n,n,n)$ calculated from eq.~\ref{qdualityidentity}.}
   
  {\proposition{The value $\gon_q(n,n,n,n)$ is equal to  the trace of the inverse of the $(n+1)$-th order quantum Hilbert matrix.}}
  
   \smallskip
Example: \\
{\scriptsize
   $H_q(3)^{-1}=\left(
\begin{array}{ccc}
 \frac{\left(q^4+q^2+1\right)^2}{q^4} & -\frac{\left(q^2+1\right) \left(q^4+1\right) \left(q^4+q^2+1\right)^2}{q^7} & \frac{\left(q^4+1\right)
   \left(q^4+q^2+1\right) \left(q^8+q^6+q^4+q^2+1\right)}{q^8} \\
 -\frac{\left(q^2+1\right) \left(q^4+1\right) \left(q^4+q^2+1\right)^2}{q^7} & \frac{\left(q^2+1\right)^4 \left(q^4+1\right)^2
   \left(q^4+q^2+1\right)}{q^{10}} & -\frac{\left(q^2+1\right) \left(q^4+1\right) \left(q^4+q^2+1\right)^2 \left(q^8+q^6+q^4+q^2+1\right)}{q^{11}} \\
 \frac{\left(q^4+1\right) \left(q^4+q^2+1\right) \left(q^8+q^6+q^4+q^2+1\right)}{q^8} & -\frac{\left(q^2+1\right) \left(q^4+1\right)
   \left(q^4+q^2+1\right)^2 \left(q^8+q^6+q^4+q^2+1\right)}{q^{11}} & \frac{\left(q^4+1\right)^2 \left(q^4+q^2+1\right)^2
   \left(q^8+q^6+q^4+q^2+1\right)}{q^{12}} \\
\end{array}
\right)$}
{\footnotesize
$Tr(H_q(3)^{-1}) = \gon_q(2,2,2,2) = \frac{q^{24}+4 q^{22}+13 q^{20}+27 q^{18}+47 q^{16}+63 q^{14}+71 q^{12}+63 q^{10}+47 q^8+27 q^6+13 q^4+4 q^2+1}{q^{12}}$
 with classical limit $\gon(2,2,2,2)= Tr(H(3)^{-1}) = 381.$}  

 Inverting an Hilbert matrix, quantum or not, takes time. In order to obtain the trace of the latter, the calculation using $\gon$ or $\gon_q$ is much faster.

\subsection{The quantum $\tet_q$ function}
\subsubsection{Quantum admissible tetrahedra}
We already defined admissibility for classical tetrahedra (see sec.~\ref{classicaladmissibletetrahedra}).
In the quantum case, and when $q$ is a root of unity,  $q = \exp(i\pi/(\ell+2))$, we also impose that the variables labelling the edges of the tetrahedron should be $q$-admissible and that the  
 four triplets associated with the four faces of a tetrahedron $T$ should be q-admissible as well (see sec.~\ref{qadmissibility}).
If the individual arguments $x$ of the function $\gon_q$ are interpreted as components of $SU(2)$  highest weights defining three irreducible representations of $SU(2)_q$ that should exist at the chosen level $\ell$, and have non vanishing $q$-dimension (then $x \leq \ell +1$),  
the $q$-admissibility condition on the triplets $(a,b,c)$,  $(b,d,f)$, $(a,e,f)$, $(c,d,e)$,  ensures that those triplets label $SU(2)$ intertwiners that exist at the chosen level.

\subsubsection{The elementary quantum $\tet_q$ function (tetrahedral function)}  
The definition of the function $\tet_q$, for an admissible tetrahedron $T$ is the same as in the classical case (see eq.~\ref{hed_definition}), but for the fact that factorials ($n!$) are replaced by q-factorials ($[n]!_q$).

\bigskip

Example. With $T= ((14, 41, 33) (50, 23, 21))$, one finds that $\tet_q(T)$ is equal to 
{\footnotesize
\[
\frac{[57]!_q}{ [3]!_q^3 [8]!_q [12]!_q [27]!_q} - \frac{ [58]!_q}{[1]!_q [2]!_q^2 [4]!_q [7]!_q [13]!_q [28]!_q} + \frac{[59]!_q}{[1]!_q^2 [2]!_q [5]!_q [6]!_q [14]!_q [29]!_q} - \frac{[60]!_q}{ [3]!_q [5]!_q [6]!_q [15]!_q [30]!_q}
\]
}
In the classical case, $q=1$, one recovers $\tet(T)=-671777611858249170324639542553600$.  \\
With $q = \exp(i \pi/60)$, the tetrahedron $T= ((14, 41, 33) (50, 23, 21))$ is $q$-admissible, and for this choice of $q$, the value of  $\tet_q(T)$, approximatively $1.53314 \times 10^{17}$,  can be exactly expressed as the largest root of the following eight degree polynomial (all its roots are real):
{\footnotesize
\begin{equation*}
\begin{split}
& -819146649600 + 22429878220800 z + 371125869811200 z^2 - 
   5461982626291200 z^3 - 69550885171932480 z^4 +  \\ &
   29181297342545280 z^5 + 1098435022151631960 z^6 - 
   153313734949066620 z^7 + z^8
\end{split}
\end{equation*}
}

\subsubsection{A topological identity of the quantum $\tet$ function: the double tetrahedron identity}
In the quantum case the two functions $\hed_{1,q}$ and $\hed_{2,q}$ are defined like  the functions $\hed_1$ and $\hed_2$  of the classical case, modulo the obvious replacement of $\tet$ by $\tet_q$ and of factorials by $q$-factorials  (see  eqs.~\ref{bipyramid1} and \ref{bipyramid2}); however in the quantum case the sign $\otimes$ appearing in the summation  $\sum _{x\in{\otimes[d,g] \cap \otimes[f, k] \cap \otimes[h,e]}}$ should be properly understood:  if q is a root of unity (\ie finite level $\ell$), there are usually less contributions to this sum over $x$ than in the classical case (\ie infinite level); this can be written explicitly in terms of $d, g$, $f,k$, $h,e$ and $\ell$.
 Without taking this precaution in the sum, the following identity would fail for small levels. Like in the classical case, one has the following analog of the quantum Biedenharn-Elliott identity: 
\begin{equation} \hed_{1,q} = \hed_{2,q}.\end{equation}

\subsubsection{The quantum $\TET_q$ function (also called $\tetrahedron_{K,q}$)}

This function is formally defined by the same expressions as in the classical case (see eq.~\ref{TET_definition}), with simple modifications: factorials should be replaced by $q$-factorials and the four triangles should be $q$-admissible.
One defines
\begin{equation} \TET_q = \frac{J_q}{E_q} \times \tet_q \label{tetToTET}\end{equation}
where $J_q$ and $E_q$ are again obtained from the classical definition of $J$ and $E$  (sec.~\ref{sec:TET}) by replacing factorials by $q$-factorials.

\subsubsection{From the quantum function $\tet_q$ to quantum $6j$ symbols}
\label{sec:TETq}

{  As in the classical case, cf.~sec.~\ref{sec:TET}, several versions of $6j$ symbols (i.e. normalized or not) can be found in the literature.
It will be enough to mention their normalized version ---that we simply call ``quantum $6j$ symbols''.}
The relation between them and the function $\TET_q$ symbols is known {  (we refer to lemma 3.11.3 of the book \cite{Carter:6J} for details).}
One has:
\begin{equation}\SixJSymbol_q((a/2, b/2, c/2), (d/2,e/2,f/2 )) = \frac{ \TET_q((a, b, c), (d,e,f))}{ N_q((a, b, c), (d,e,f))}\end{equation}
where 
\begin{equation}N_q((a, b, c), (d,e,f)) =  \sqrt{ |\ThetaGraph_{K,q}(a, b, c) \ThetaGraph_{K,q}(b,d, f) \ThetaGraph_{K,q}(a,e,f) \ThetaGraph_{K,q}(c,d,e)| }\end{equation}
{  Then, using \ref{tetToTET} and the relation between $\ThetaGraph_{K,q}$ and $\gon_q$, one obtains:}
{\proposition
\begin{equation}
\SixJSymbol_q((a/2, b/2, c/2), (d/2, e/2, f/2)) = \frac{\tet_q((a, b, c), (d, e, f))} { \sqrt{\vert \gon_q(a,b,c) \gon_q(b,d,f) \gon_q(a,e,f) \gon_q(c,d,e)  \vert }}
\label{hedfromq6j} 
 \end{equation}}    

\subsubsection{Other properties}
The other properties of the classical function $\tet$ described in the first part of this paper have a quantum counterpart. 
The writing of those properties in terms of q-functions is left to the reader.

\section{Miscellaneous}

\subsection{Evaluation in spin networks: a very short summary}
Quantities called triangular functions (or theta symbols), and $6j$ symbols, show up in many branches of physics or mathematics where they are usually defined as $\ThetaGraph_K$, and as ${\tetrahedron}_U$ (or  $\SixJSymbol$).
Existence of their integer-valued partners $\gon$ and $\tet$ is usually not even mentioned but it remains that the definition of these two functions fits naturally in the theory of spin networks.
We give below a very short summary of the evaluation procedures in this theory in order to ease the comparison between functions and concepts defined there and the two functions $\gon$ and $\tet$   ---we write ``procedures'' with a plural because very often in the literature, only one procedure is mentioned (they are rare exceptions, for instance \cite{GaroufalidisEtAl}).  

\paragraph{Classical spin networks.}
\label{sec:classicalspinnetworks}

Let $\Gamma$ be a tri-valent graph and a {\sl coloring} $\gamma$ of its edges by natural numbers ($0$ is allowed). 
$\Gamma$ is assumed to be admissible:  at each vertex of $\Gamma$ the triplet of integers should be admissible (in particular their sum is even).
There are closed graphs (no {  external lines}) and open graphs. {  A spin network is defined by a pair $(\Gamma, \gamma)$, for instance ($\ThetaGraph$, $\gamma$) or  (${\tetrahedron}$, $\gamma$)}.

\smallskip
{\small
{\sl Warning}: 
Starting from the spin network defined by the labelled (\ie colored) theta graph ($\ThetaGraph$, $\gamma$) where admissibility is imposed on the three edges meeting at each vertex, one obtains by duality an admissible triangle, like in section \ref{admissibletriple}, 
where admissibility is defined around a triangular face.  The tetrahedron is a self-dual figure, but duality permutes the notions of admissibility:
the tetrahedron spin-network defined by the tri-valent graph (${\tetrahedron}$, $\gamma$) is, by definition, admissible (as a spin network), but the tetrahedron obtained from the labels defined by ${\gamma}$ (see fig.~\ref{dualtetrahedronT}) is usually ${\sl not}$ admissible in the sense of section \ref{classicaladmissibletetrahedra},  however its dual, obtained by permuting the two triplets in the notation $((a, b, c), (d, e, f))$, is.}
\smallskip
 
Any spin network ($\Gamma$, $\gamma$) can be {\sl evaluated}. {  
The meaning of the obtained number depends on the theory (physical or mathematical) in which spin networks are used: in the original Penrose theory \cite{Penrose}, which was a simple model for a quantum space, it was used to calculate the probabilities of various spin values;
in various theories of quantum gravity this number is interpreted as an amplitude for a quantum state of space-time itself; other interpretations arise in knot theory, or in the geometry of $3$-manifolds.}
An {\sl evaluation procedure} is defined for each colored connected tri-valent graph by the following set of rules : 
\begin{itemize}
\item A value is set for the loop graph: see below.
\item A value is set for the theta graph $\ThetaGraph$: see below.
\item A value is set for the tetrahedron graph $\tetrahedron$: see below.
\item Three rules (called recoupling rules) allow one to simplify the evaluation of $\Gamma$ on the coloring $\gamma$ by replacing a given graph by a sum of products (or ratios) of numbers, which involves only evaluations of the loop graph, the theta graph, the tetrahedron graph, and, if the original graph was open, a family  of left over graphs which are $3$-valent trees or straight lines. As we do not need these rules in the present paper, we refer the reader to the literature on the subject, for instance \cite{GaroufalidisEtAl}.
\end{itemize}

There are several possible evaluation procedures (at least four kinds). We call them {\sl Integer evaluation} $\Gamma_\Z(\gamma)$, {\sl Penrose evaluation} $\Gamma_P(\gamma)$, {\sl Kauffman evaluation} $\Gamma_K(\gamma)$, and {\sl Unitary evaluation} $\Gamma_U(\gamma)$. They differ by normalization prescriptions (see below).

For a given coloring $\gamma$ of a tri-valent graph $\Gamma$ we call $v_1, v_2, v_3$ the values of the three edges emanating from the vertex $v$ of $\Gamma$. Call $V(\Gamma)$ the set of vertices,  $E(\Gamma)$ the set of edges, and $\gamma(e)$ the coloring of the edge $e$. 
Define $\sigma_v = \frac{v_1+v_2+v_3}{2}$ and $\Theta_v = (-1)^{\sigma_v} \, (\sigma_v+1) \, M(\sigma_v - v_1, \sigma_v - v_2, \sigma_v - v_3)$ where $M$ denotes the multinomial coefficient. Call 
\begin{equation}
\begin{split}
{\mathcal J}(\Gamma) &= \Pi_{v\in V(\Gamma)} \; (\sigma_v-v_1)! (\sigma_v-v_2)!(\sigma_v-v_3)!,\\
{\mathcal E}(\Gamma) &= \Pi_{e\in E(\Gamma)} \; \gamma(e) !  \\
{\mathcal N}(\Gamma) &= \Pi_{v\in V(\Gamma)} \; \sqrt{\vert \Theta_v \vert}.
\end{split}
\end{equation}

For all evaluation procedures, the evaluation of the loop graph on the integer $n$ is $(-1)^n (n+1)$.\\
For a spin network ($\Gamma$, $\gamma$) with integer evaluation $\Gamma_\Z(\gamma)$, one defines the other evaluation procedures as follows: \\
\begin{equation}
\Gamma_P(\gamma) = {\mathcal J}(\Gamma) \times\Gamma_\Z(\gamma), \quad \Gamma_K(\gamma)= \frac{ {\mathcal J}(\Gamma)}{{\mathcal E}(\Gamma)}  \times\Gamma_\Z(\gamma), \quad \Gamma_U(\gamma) =  \frac{1}{{\mathcal N}(\Gamma)}  \times\Gamma_\Z(\gamma).
\label{evaluationProcedures}
\end{equation}
It is therefore enough to specify the integer evaluations $\Gamma_\Z(\gamma)$  where $\Gamma$ is either the  theta graph $\ThetaGraph$ (dual to the triangle with edges $(v_1, v_2, v_3)$),  or the tetrahedron  graph $\tetrahedron$, with coloring 
dual to  the admissible tetrahedron $T$ with labels $((a,b,c),(d,e,f))$.
For these two graphs we set: 
\begin{equation}
\begin{split}
{\ThetaGraph}_\Z(\gamma) =& \,  \Theta_v  =  (-1)^\sigma \, (\sigma+1) \, M(\sigma - v_1, \sigma - v_2, \sigma - v_3) = (-1)^\sigma \,  \gon(v_1, v_2, v_3),\\
{\tetrahedron}_\Z(\gamma)  =&  \tet(T) = \tet((a,b,c),(d,e,f)).
\end{split}
\end{equation}
The function $\gon$ was defined in eq.~\ref{gondefinition}, and the function $\tet$  in eq.~\ref{hed_definition}.   

\smallskip

For the theta graph, {  using \ref{evaluationProcedures} and setting} 
\begin{equation}
{\mathcal J}({\ThetaGraph}) = ((\sigma-v_1)!  (\sigma-v_2)!  (\sigma-v_3)!)^2,  \qquad {\mathcal E}({\ThetaGraph})=v_1! v_2! v_3!, \qquad {\mathcal N}({\ThetaGraph})= {\vert \Theta_v \vert},
\end{equation}
one finds:
\begin{equation}
\begin{split}
{\ThetaGraph}_P(\gamma) =&  (-1)^\sigma (\sigma+1)! (\sigma -v_1)!  (\sigma_v-v_2)!(\sigma_v-v_3)!, \\
{\ThetaGraph}_K(\gamma) =&  (-1)^\sigma (\sigma+1)!  \frac{ (\sigma_v-v_1)! (\sigma_v-v_2)!(\sigma_v-v_3)!}{v_1! v_2! v_3!}, \\
{\ThetaGraph}_U(\gamma) =&  (-1)^\sigma.
\end{split}
\end{equation}
For the tetrahedron graph,  calling $A,B,C,D$ its vertices {  and using \ref{evaluationProcedures}} , one finds:
\begin{equation}
\begin{split}
{\tetrahedron}_P(\gamma) =& {\mathcal J}(\tetrahedron) \, \tet((a,b,c),(d,e,f)) \,  \text{with} \,  {\mathcal J}(\tetrahedron) = \Pi_{v\in\{A,B,C,D\}} (\sigma_v - v_1)! (\sigma_v - v_2)! (\sigma_v - v_3)!  ,\\
{\tetrahedron}_K(\gamma) =& \TET((a,b,c),(d,e,f)),  \,   \text{with}  \, \TET \,  \text{as in eq.~\ref{TET_definition}},\\
{\tetrahedron}_U(\gamma) =& \SixJSymbol((a/2, b/2, c/2), (d/2, e/2, f/2)),  \text{see also eq.~\ref{tetfrom6j}}.
\end{split}
\end{equation}

 \begin{figure}[htb]
\begin{center}
\includegraphics[width=0.3\textwidth]{tetrahedronT.pdf}
\includegraphics[width=0.3\textwidth]{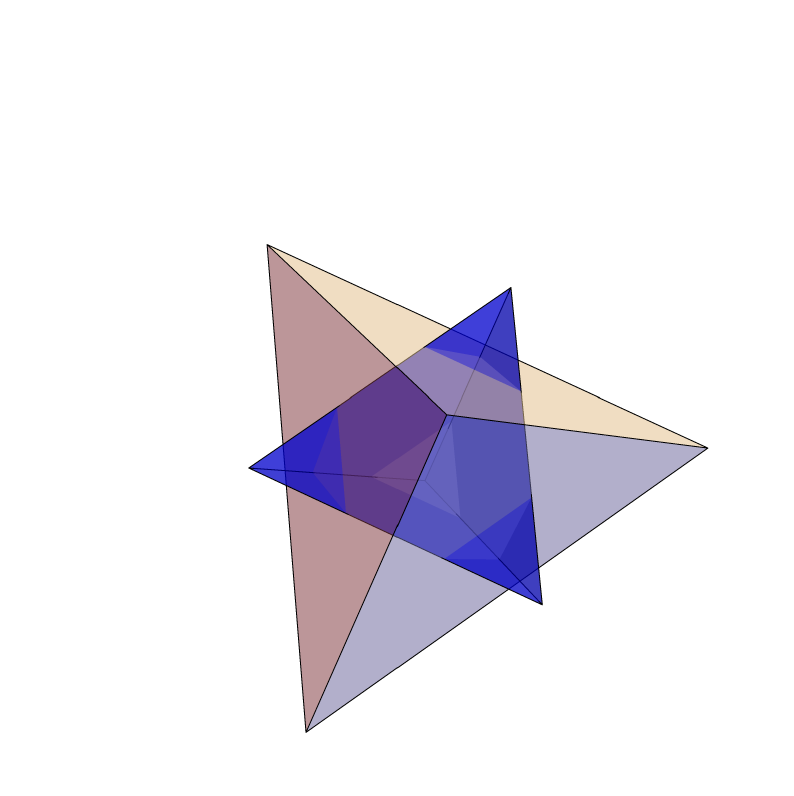}
\includegraphics[width=0.3\textwidth]{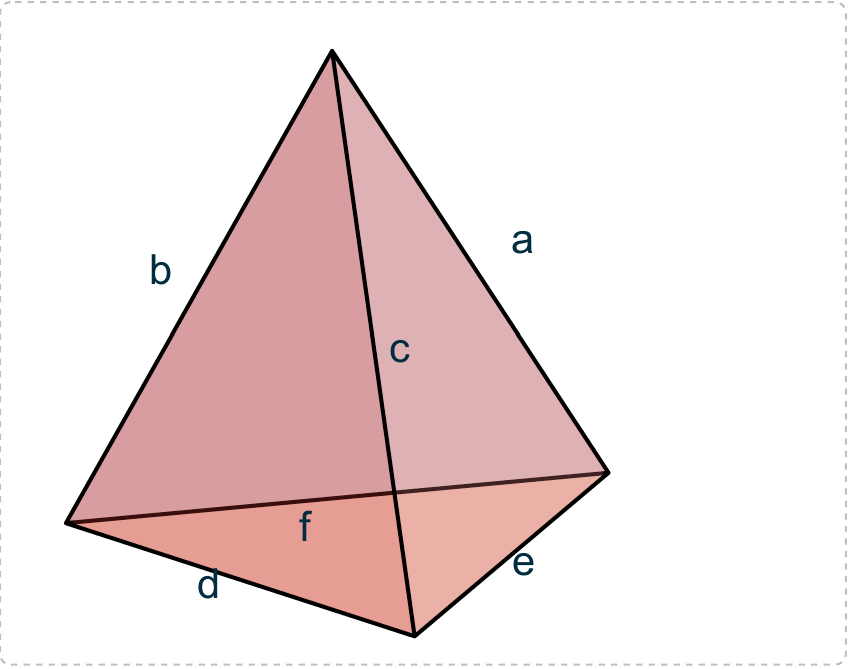}
\caption{The tetrahedron of fig.~\ref{tetrahedronT} and the tetrahedron with a dual labelling.} 
\label{dualtetrahedronT}
\end{center}
\end{figure}

\paragraph{Quantum spin networks.}
By replacing integers by $q$-integers, one can develop a quantum counterpart of the theory of classical spin networks. We refer the reader to the abundant literature on the subject.

\subsection{About integral tetrahedra}

Consider a {\sl labelled tetrahedron} as given by fig.~\ref{tetrahedronT}. 
We called {\sl admissible tetrahedron} a labelled tetrahedron such that its faces are admissible triplets (in the sense defined in section \ref{admissibletriple}). The labels ${a,b,c,d,e,f}$ are therefore such that:
$(1)$ They are non-negative integers, $(2)$ They obey triangular inequalities such that the four faces are indeed triangles (which are possibly degenerate, and possibly flat), $(3)$ The perimeters of the four triangular faces are even.

An admissible tetrahedron $T$, and more generally a labelled tetrahedron obeying at least the above condition $(2)$, is called Euclidean, Minkowskian, or flat, if its Cayley-Menger determinant $\Delta$  (\ref{CayleyMenger}) is positive, negative, or zero.
So far we did not impose any constraint on the sign of $\Delta$, so the admissible tetrahedra can be of any of those three types. 
$T$ is called degenerate if one or several edges vanish.
\begin{equation}
\label{CayleyMenger}
\Delta =  \text{det} \, \left(
\begin{array}{ccccc}
 0 & a^2 & c^2 & e^2 & 1 \\
 a^2 & 0 & b^2 & f^2 & 1 \\
 c^2 & b^2 & 0 & d^2 & 1 \\
 e^2 & f^2 & d^2 & 0 & 1 \\
 1 & 1 & 1 & 1 & 0 \\
\end{array}
\right)
\end{equation}

 If an admissible tetrahedron is Euclidean, which implies that it has a positive volume $V = \sqrt{\Delta/ 288}$  and that it defines (up to isometry) a metric tetrahedron that one can embed in $\R^3$, it  defines an Euclidean integer-sided tetrahedron.
 Notice that the number of non-congruent such tetrahedra (let us say with given largest side) is smaller than the number given by the OEIS integer sequence A097125, although this sequence counts the number of non-congruent Euclidean integer-sided tetrahedra with largest side $n$, because the latter are not necessarily even (an integer-sided tetrahedron obeying the condition $(3)$ is called even) whereas ours are.
 Notice also that an admissible tetrahedron of Euclidean type has no reason to be Heronian in general (we do not impose integrality of surface and volume).
 
 \subsection{Hom-spaces.}
\label{Hom-spaces}
{  A linear map between two representation spaces that commutes with the group action is, by definition, an equivariant linear map between these two representations; it is also called an intertwiner. In other words, an intertwiner is a homomorphism of representation spaces.
Let $a$ and $b$ denote two finite dimensional unitary irreducible representations of some compact group.
Their tensor product $a \otimes b$ is a (usually not irreducible) representation, 
its decomposition into irreducible representations reads $a \otimes b = \bigoplus_i c_i$ where each of the $c_i$ appearing in this direct sum
is an irreducible subrepresentation of $a \otimes b$. 
There may be repetitions of the $c_i$ that occur in that direct sum (multiplicity).}
 If the chosen group is the Lie group $SU(2)$, the multiplicity is always equal to $1$, this is a particular feature of $SU(2)$.
{  In terms of highest weights of irreps, the fact that $c$ should be such that $\vert a - b \vert \leq c \leq \vert a + b \vert$ (the law of ``coupling of spins'')  makes the triplet $(a,b,c)$ admissible and translates as a usual triangle inequality.
In terms of representations spaces, the inclusions of these irreducible subrepresentations are intertwiners maps; the adjoint $\gamma$ of such an inclusion $\gamma^\star$ is a surjective map, so we also have surjective intertwiners from $V_a \otimes V_b$ to $V_{c_i}$.
The composition $\gamma \gamma^\star$ is the identity on $V_{c_i}$ whereas the composition  $\gamma^\star \gamma$ is a projector defined in the vector space $V_a\otimes V_b$.
An intertwiner from $a \otimes b$  to $c$ has a pictorial description as a triangle, with edges labelled by the irreducible representations $a$, $b$ and $c$.
These edges are a priori oriented, however Frobenius reciprocity tells us that one can replace a homomorphism {\sl to} a representation by a homomorphism {\sl from} its conjugate representation. 
In the case of $SU(2)$, an irreducible representation is equivalent to its conjugate so one may forget about orientation of the edges of the triangle $a,b,c$ and picture intertwiners as triangles with unoriented edges. The above discussion is of course standard.}
The triangle $(a,b,c)$ is a convenient visualization of the space of intertwiners $Hom_{SU(2)} (V_a \otimes V_b \otimes V_c, \C)$, which is $1$-dimensional.
{   Intuitively,} one can think of this admissible triangle as describing a particular kind of ``relation'' between three irreducible representations of $SU(2)$ and one can also think of an admissible tetrahedron as describing ``relations between relations'', where the four faces of the tetrahedron refer to four compatible intertwiners.

\subsection{From $SU(2)$ to $SU(N)$, or to $G$.}
The $SU(2)$ group (classical or quantum) underlies the various constructions or considerations described previously. 
For instance, from the very beginning, the edges of the triangles, or of the tetrahedra, are labelled by non-negative integers referring to highest weights of irreducible representations of $SU(2)$. 
If one replaces the latter by another simple (or semi-simple) Lie group $G$, of higher rank, one can of course look for possible generalizations and it is clear that such generalizations should ``exist''.
The function $\gon(a,b,c)$, classical or quantum, can be seen as a generalization from the $1$-simplex to the $2$-simplex of the function giving the dimension of an $SU(2)$ irreducible representation (the function $\dimension(a)=a+1 \in \N$), and the $\tet$ function, classical or quantum, as a generalization of the latter to the $3$-simplex.
The dimension function $\dimension$ is explicitly known for all irreps of simple Lie groups $G$, but things get tricky already at the level of the triangle function $\gon$ when $G$ is not $SU(2)$. 
Indeed, apart from the fact that one can no longer, in general, identify a representation with its conjugate, the spaces of intertwiners $Hom_{G} (V_a \otimes V_b \otimes V_c, \C)$, equivalently $Hom_{G} (V_a \otimes V_b, \overline{V_c})$, are no longer $1$-dimensional:  multiplicities appear ---they are given by the Littlewood-Richardson coefficients in the $SU(N)$ case.
In other words, triangles spaces (Hom-spaces) are not fully specified by a compatible choices of  three irreducible representations. 
From a combinatorial point of view,  if $G=SU(N)$ and if the space of intertwiners (for given $a,b,c$) is of dimension $s$,  
one can draw $s$ distinct pictographs (one can use honeycombs, or, equivalently,  O-blades or  Berenstein-Zelevinsky diagrams),  with three {\sl external} sides labelled by the same triplet $(a,b,c)$ of irreps but differing by their ``inner'' contents.
In the $G=SU(2)$ case one can trade the three ``external edge variables'' $(a,b,c)$ for three ``internal edge variables $(m,n,p)$ as we did in the first section. 
In the rank $2$ case, taking $G=SU(3)$, there are $2\times 3=6$ numerical external edge variables (the components of the three highest weights $a,b,c$ that one can attach to three copies of an $A_2$ Dynkin diagram) but the intertwiners themselves are labelled by $8$ non-negative ---and non-independent--- integer parameters; using a syzygie relation one can get rid of one non-negative parameter and stay with $7$ independent (but not all necessarily positive) parameters.  
In this $SU(3)$ case the argument of the $\gon$ function (or of the $\ThetaGraph_K$ function in another normalization) cannot be specified by the triplet of irreps $(a,b,c)$ alone, because this argument should also specify the choice of a pictograph (a honeycomb for instance) whose description usually requires one more parameter.
Even in the relatively simple situation of $SU(3)$ the $\gon$ function is not known explicitly in full generality although several people have worked out explicit expressions for it, using one or another normalization (G. Kuperberg and his student D. Kim (2003), see \cite{Kim}, 
A. Ocneanu and his student L. Suciu (1997), see \cite{Suciu}), but this was done only for particular families of irreps $a,b,c$.
In the $SU(4)$ case (where one needs $3\times 3+3=12$ independent parameters, so $3$ more than the components of the three given highest weights), and more generally for $SU(N)$, $N > 3$, or for other simple Lie groups, we are not aware of any general explicit result for the triangle function. 
As for the tetrahedral function (we remind the reader that for $G=\SU(2)$, and in the unitary normalization, it is the usual $6j$ symbol), the four faces of the tetrahedron should be labelled by appropriate compatible honeycombs fully specifying the triangle spaces under consideration, but even for $SU(3)$, and apart from very few special cases (low dimension of irreps and no multiplicity) worked out by physicists at the time of the eightfold way, we are not aware of any published general result.

\subsubsection*{{  Appendix: About Clebsch-Gordan coefficients, $3j$ and $6j$ symbols}}
\label{appendix}
{ 

The name ``Clebsch-Gordan coefficients'' comes from two 19-th century German mathematicians, A. Clebsch and P. Gordan, who worked on problems of invariant theory.
These numbers, in physics, arise in the quantum treatment of angular momentum: the coupling of two spins.
In a more general setting,  let  $V_j$ be the vector space of an irreducible and unitary representation (irreps) of some compact Lie group,  
the tensor product of two irreps can be decomposed as a sum of irreducible components.
An intertwiner being chosen (an equivariant morphism mapping $V_{j_1} \otimes V_{j_2}$ to some chosen $V_J$, see the discussion on Hom-spaces in section \ref{Hom-spaces}), 
one can evaluate this morphism on the tensor product of two vectors $m_1 \in V_{j_1}$, $m_2 \in V_{j_2}$ and take the inner product of its image with  vectors $M \in V_J$, one obtains a number.
We now assume that the group is $SU(2)$. In that case, and for three chosen irreps $V_{j_1}$, $V_{j_2}$, $V_{J}$,  such an intertwiner is unique up to a scaling factor (no multiplicity); moreover the labels $j_i$, or $J$ can be chosen as half-integers (spin variables).
For vectors $m_j$, or $M$, belonging to an appropriate normalized basis in the representation spaces, the numbers obtained as before are called Clebsch-Gordan coefficients and denoted $\text{ClebschGordan}[{j_1, m_1}, {j_2, m_2}, {J, M}]$ or  $\langle {j_1},\, {j_2};\; {m_1},\, {m_2}| J,\, M\rangle$.
Their properties are described in many textbooks, review articles, and encyclopedias.
Equivalently one can think of the evaluation as being defined on the vector space $V_{j_1} \otimes V_{j_2} \otimes V_J$, this leads to coefficients called Wigner $3j$ symbols that are traditionally written using round braces, as below, and look more symmetrical than the Clebsch-Gordan coefficients;
the relation between both is as follows:} 
  \begin{equation}
  \label{3jtoCG}
\langle {j_1},\, {j_2};\; {m_1},\, {m_2}| J,\, M\rangle =\sqrt{2 J+1} (-1)^{{j_1}-{j_2}+M} \Bigg(\medspace 
\begin{array}{ccc}
 {j_1} & {j_2} & J \\
 {m_1} & {m_2} & -M \\
\end{array}
\medspace \Bigg)
  \end{equation}
{  The signifiance of Clebsh-Gordan coefficients in quantum mechanics became clear after the work of Wigner (1927, in German), see the bibliography in \cite{BiedenharnLouck};  properties and applications of these coefficients, and of $3j$ symbols, were then studied in many articles and textbooks (among classical textbooks one should certainly mention \cite{Wigner} and \cite{Messiah}).}

\bigskip 
{ 
The $6j$ symbols were introduced by E.P. Wigner in 1940 in his study of the coupling of three angular momenta (spins); this work was published much later (\cite{Wigner1965}, 1965),  but the usefulness of these symbols was immediately recognized and their properties discussed in articles and textbooks (again we can mention \cite{Messiah}, 1962). It was also observed that the Wigner $6j$ symbols were simply related to another family of coefficients, the Racah coefficients, that had been defined, differently, by Racah \cite{Racah}, in 1942 (see below).
The Wigner $6j$ symbols are often defined as a sum over a product of four $3j$ symbols (or four Clebsch-Gordan coefficients), see eq.~\ref{6jfrom3j}.
A standard notation for $6j$'s uses curly braces, as on the left hand side of equation \ref{6jfrom3j}. 

The Racah coefficients,  introduced in 1942, \cite{Racah}, were defined as a formally infinite sum but for given arguments, only finitely many terms of this sum are nonzero.
We do not use this formula in this paper but the interested reader is referred for instance to \cite{Racahcoefficient} and references therein.
As already mentioned, it was shown long ago that the coefficients defined by the Racah formula are equal (up to a possible sign) to the Wigner $6j$ symbols.

Warnings:\\
$\bullet$ Another family of coefficients, the so-called recoupling coefficients, which are sometimes called unitary Racah symbols, frequently appear in the literature, see the discussion after eq.~\ref{recouplingfrom6j}.\\
$\bullet$
Still another variant of the $6j$ symbols, also called by the same name, and  sometimes also denoted by curly braces, but differing from the Wigner $6j$'s  by normalization factors, 
can often be found, mostly in the mathematical literature.
In the few places where both kinds of symbols are used simultaneously, the authors usually call ``normalized $6j$ symbols''  the Wigner ones, i.e. those that are given by eqs.~\ref{TETto6j} or \ref{6jfrom3j},
although, for other people (and for us), the latter are just``$6j$ symbols''.
This is a confusing issue which is discussed at the beginning of section \ref{sec:6jfromtet}.

\bigskip
Although we do not use in this paper the formula expressing Wigner $6j$ symbols in terms of $3j$ symbols, we display it below, for the convenience of the reader.
}
{\tiny 
\begin{equation}
\begin{split}
\Bigg\{\medspace 
\begin{array}{ccc}
 {j_1} & {j_2} & {j_3} \\
 {j_4} & {j_5} & {j_6} \\
\end{array}
\medspace \Bigg\} &= \sum_{ { }^{{m_k}=-{j_k}}_{k=1,\ldots 6}}^{j_k}  (-1)^{\sum_{k=1}^6(j_k-m_k)}
 \Bigg(\medspace 
\begin{array}{ccc}
 {j_1} & {j_2} & {j_3} \\
 -{m_1} & -{m_2} & -m_3
\end{array}
\medspace \Bigg) \Bigg(\medspace 
\begin{array}{ccc}
 {j_4} & {j_2} & {j_6} \\
 {m_4} & {m_2} & -{m_6} \\
\end{array}
\medspace \Bigg)
 \Bigg(\medspace 
\begin{array}{ccc}
 {j_1} & {j_5} & {j_6} \\
 {m_1} & - m_5&m_6 \\
\end{array}
\medspace \Bigg) \Bigg(\medspace 
\begin{array}{ccc}
 {j_4} & {j_5} & {j_3} \\
 -{m_4} & {m_5} & {m_3} \\
\end{array}
\medspace \Bigg) 
\end{split}
\label{6jfrom3j}
\end{equation}
}
Because of Kronecker constraints coming from the $3j$ symbol, {$m_3=-(m_1+m_2)$, $m_6=(m_2+m_4)$, $m_5=m_1+m_2+m_4$}, the previous sum can be written as a sum over three indices only:
{\tiny 
\begin{equation}
\begin{split}
&\Bigg\{\medspace 
\begin{array}{ccc}
 {j_1} & {j_2} & {j_3} \\
 {j_4} & {j_5} & {j_6} \\
\end{array}
\medspace \Bigg\}=\sum_{{m_1}=-{j_1}}^{{j_1}} \sum_{{m_2}=-{j_2}}^{{j_2}} \sum _{{m_4}=-{j_4}}^{{j_4}} \; (-1)^{{j_1}+{j_2}+{j_3}+{j_4}+{j_5}+{j_6}-{m_1}-2 {m_2}-3 {m_4}} 
\\
&
 \Bigg(\medspace 
\begin{array}{ccc}
 {j_1} & {j_2} & {j_3} \\
 -{m_1} & -{m_2} & {m_1}+{m_2} 
\end{array}
\medspace \Bigg) \Bigg(\medspace 
\begin{array}{ccc}
 {j_4} & {j_2} & {j_6} \\
 {m_4} & {m_2} & -{m_2}-{m_4} \\
\end{array}
\medspace \Bigg)
 \Bigg(\medspace 
\begin{array}{ccc}
 {j_1} & {j_5} & {j_6} \\
 {m_1} & -{m_1}-{m_2}-{m_4} & {m_2}+{m_4} \\
\end{array}
\medspace \Bigg) \Bigg(\medspace 
\begin{array}{ccc}
 {j_4} & {j_5} & {j_3} \\
 -{m_4} & {m_1}+{m_2}+{m_4} & -{m_1}-{m_2} \\
\end{array}
\medspace \Bigg) 
\end{split}
\tag{\ref{6jfrom3j}$'$}
 \label{sec:6jfrom3jprime}
\end{equation}
}

Equivalently, {  in terms of Clebsch-Gordan coefficients},

{\tiny 
\begin{equation}
\begin{split}
&
\Bigg\{\medspace 
\begin{array}{ccc}
 {j_1} & {j_2} & {j_3} \\
 {j_4} & {j_5} & {j_6} \\
\end{array}
\medspace \Bigg\}
= \sum _{{m_1}=-{j_1}}^{{j_1}} \sum _{{m_2}=-{j_2}}^{{j_2}} \sum _{{m_4}=-{j_4}}^{{j_4}}
  \frac {(-1)^{{j_1}-{j_2}+{j_3}+{j_4}-{j_5}+{j_6}-{m_1}-{m_4}}} {(2 {j_3}+1) (2 {j_6}+1)} 
    \langle {j_1}, \medspace {j_2} ; \medspace -{m_1},\medspace
   -{m_2}\medspace |\medspace {j_3},\medspace -{m_1}-{m_2}\rangle \\
   & \langle {j_4},\medspace
   {j_2};\medspace {m_4},\medspace {m_2}\medspace | \medspace {j_6},\medspace {m_2}+{m_4}\rangle
   \langle {j_1},\medspace {j_5};\medspace {m_1},\medspace -{m_1}-{m_2}-{m_4}\medspace | \medspace
   {j_6},\medspace -{m_2}-{m_4}\rangle \langle {j_4},\medspace {j_5};\medspace -{m_4},\medspace {m_1}+{m_2}+{m_4}   \medspace |  \medspace {j_3},\medspace {m_1}+{m_2}\rangle
   \end{split}
   \tag{\ref{6jfrom3j}$''$}
 \label{sec:6jfrom3jdbleprime}
\end{equation}
}

{  In order to save space, and also because it is used in the program Mathematica, the display notation for $6j$'s using curly braces is often replaced, in the main body of the present article, by the in-line  $\SixJSymbol$ notation, defined as follows:}

\begin{equation}
\SixJSymbol((j_1,j_2,j_3),(j_4,j_5,j_6))=
\Bigg\{
\begin{array}{ccc}
 {j_1} & {j_2} & {j_3} \\
 {j_4} & {j_5} & {j_6} \\
\end{array}
\Bigg\}
\label{MathematicaNotation6J}
\end{equation}

\ommit{
\begin{equation}
W(j_1,j_3,j_6,j_4;j_2,j_5)=
(-1)^{j_1+j_3+j_6+j_4}
\Bigg\{
\begin{array}{ccc}
 {j_1} & {j_2} & {j_3} \\
 {j_4} & {j_5} & {j_6} \\
\end{array}
\Bigg\}
\label{RacahNotationW}
\end{equation}
}

\subsection*{Acknowledgments}
We acknowledge many conversations with A. Ocneanu on the geometrical structures underlying the generalizations of  $6j$ symbols to simple Lie groups of higher ranks, 
and more generally on ``higher representation theory'', although none of these ``higher'' constructions is described in the present paper.
We also thank O. Ogievietsky who suggested that one should look for elliptic generalizations of the functions $\gon$ and $\tet$, some preliminary results being promising, but they are not mentioned in the present paper either.
{  The author thanks the anonymous referee for his careful reading of the manuscript and for the many suggestions which helped to improve the clarity of the text.}
The author also acknowledges Invitational Fellowship for Research in Japan of the Japan Society for the Promotion of Science which supported his stay at the University of Tokyo where this work was partially conducted.

{}

 \end{document}